\documentclass[
preprint,
 amsmath,amssymb,
 aps,
pra,
]{revtex4-1}

\usepackage{graphicx}
\usepackage{dcolumn}
\usepackage{bm}
\usepackage[utf8]{inputenc}
\usepackage{xcolor}

\newcommand*\n[1]{\underline{#1}}

\begin{document}


\title{Convexity properties of superpositions of degenerate bipartite eigenstates}

\author{Natalia A. Giovenale}
\email{ngiovenale@famaf.unc.edu.ar}
\author{Federico M. Pont}%
 \email{pont@famaf.unc.edu.ar}

\author{Pablo Serra}
\email{serra@famaf.unc.edu.ar}

\author{Omar Osenda}
\email{osenda@famaf.unc.edu.ar}
\affiliation{ Facultad de Matem\'atica, Astronom\'{\i}a, F\'{\i}sica y Computaci\'on, Universidad Nacional de C\'ordoba y Instituto de F\'{\i}sica Enrique Gaviola - CONICET,  Av. Medina Allende s/n, Ciudad Universitaria, CP:X5000HUA C\'ordoba, Argentina,}


\date{\today}

\begin{abstract}
The entanglement content of superpositions of pairs of degenerate eigenstates 
of a bipartite system are considered in the case that both are also eigenstates 
of the $z$ component of the total angular momentum. It is shown that the von 
Neumann entropy of the state that is obtained tracing out one of the parts of 
the system has a definite  convexity (concavity) as a function of the 
superposition parameter and that its  convexity (concavity) can be predicted 
using a quantity of information that measures the entropy shared by the states 
at the extremes of the superposition. Several  examples of two particle 
system, whose eigenfunctions and density matrices can be obtained 
exactly, are analyzed thoroughly.

\end{abstract}

\maketitle


\section{Introduction}\label{sec:level1}

As the old recipe goes, the only thing that someone needs to prepare a pure 
state with 
well defined quantum numbers is a complete set of compatible observables of the 
quantum system whose  states are to be prepared. Then, applying a given sequence 
of one-dimensional projectors to an arbitrary pure state, it is possible to 
label 
the resulting state with the quantum numbers associated to each projector. The 
sequence contains only one projector for each observable which is taken from the 
corresponding spectral decomposition. Only if the observable set is complete the 
state at the end of the sequence will be an element of the basis that expands 
the Hilbert space of the system and, consequently, can not be written as a 
superposition of states generated following the recipe but with different 
projectors. Of course, one-dimensional projectors are elusive objects to be 
actually constructed, in particular when the observable has a continuum 
spectrum.

The drive to process quantum information, to control the states where it is 
stored 
and to palliate the unwanted effects of decoherence mechanisms has resulted in a 
host of methods to produce pure or almost pure states in a reliable and 
repetitive way in different quantum systems.  A short list of examples includes, 
effective pure states in Magnetic Nuclear Resonance 
\cite{Gershenfeld1997,Cory1997,Vandersypen2001}, state preparation of coupled 
electron states in quantum dots \cite{Petta2005} and synthesis of arbitrary 
states in superconducting qubits \cite{Hofheinz2009}. Moreover, the 
availability of quantum states made of specific superpositions of pure states or
measurement-prepared states
could improve the means to perform a given quantum information task 
\cite{Knill2001,Abrams1999,Obrien2009,lofranco}.

Some years ago, Vaziri, Weihs and Zeilinger \cite{Vaziri2002} showed that for a 
particular Quantum Information task, Quantum Cryptography, the availability of 
photon states made of superpositions of orbital angular momentum eigenstates 
resulted in an expanded alphabet to be used to code information, with 
possibilities beyond the simpler alphabet formed by the two polarization states. 
In Reference \cite{Vaziri2002}  superpositions between Gaussian and 
Laguerre-Gaussian states were considered, but over the years numerous other 
examples of states that are composed of different orbital angular momentum 
states have been analyzed including photon pairs with
entangled orbital angular momentum \cite{Jack2009}, 
composite Laguerre-Gaussian beams with tunable intensity and phase distribution 
\cite{Parisi2014} and elliptic Gaussian optical vortices \cite{Kotlyar2017}. 
Besides, it has been shown experimentally that photonic states with large 
orbital angular momentum, often called qudits, can be cycled among them using 
standard optical elements \cite{Schlederer2016}.

In other physical systems, to our knowledge, there are
neither the requirement of an specific superposition of
orbital angular momentum states to perform a task, nor
a protocol to achieve it, as it is for the neat example
for photons presented above. Anyway, this scenario is changing because of the 
appearance of experiments and theoretical proposals in which the properties of a 
photonic state are transferred to a condensate system \cite{Dall2014}. 

Nevertheless, there are at least two reasons to consider superpositions of degenerate states. One comes 
from the properties of many-body eigenstates, and another one arises from considerations about 
the Hilbert space expanded by all the superpositions that can be formed using a 
set of pure states. When considering
many-body models the eigenstates of the Hamiltonian
are, most commonly, degenerate and depending on the problem that is under study 
the entanglement content could be calculated for a mixture or for a superposition of them. This subtle point was early 
acknowledged by Osborne and Nielsen \cite{Osborne2002}. They were studying the 
behaviour of the entanglement in the quantum phase transition that appears in 
the transverse Ising model and discussed the differences in entanglement 
content between the low temperature limit of the Gibbs state, that is a equally 
weighted mixture of the two degenerate ground states that the system admits, 
and the entanglement of one of those ground states.

 Since the work of Osborne and Nielsen~\cite{Osborne2002}, the entanglement 
content of the degenerate eigenstates of many-body problems has been addressed 
to asses the relationship between long-range interactions and the entanglement 
of spin pairs separated at different lengths \cite{Gaudiano2008}, the 
relationship between entanglement and symmetry in permutation-symmetric states 
\cite{Markham2011}, the difference between symmetrical superpositions of ground 
states and symmetry breaking ones  using mutual information \cite{Hamma2016}, the entanglement properties of the whole set of eigenstates of different 
Hamiltonians \cite{Vidmar2017a,Vidmar2017b}. For instance, in Reference~\cite{Gaudiano2008} the entanglement was calculated for equally weighted 
mixtures of degenerate eigenstates, while Markham in Reference~\cite{Markham2011} 
considers superpositions of states where its coefficients are 
given by complex numbers with modulus equal to one, so the normalization 
constant of the superposition is equal to the number of states that enters in 
it.

On the other hand, when all the superpositions of a set of $N$ 
eigenstates are considered, since they are contained in a 
compact space, a given continuous functional of the states should reach a set 
of extrema. So, any given entanglement measure will reach a
number of extrema. But, for which superpositions are 
those extrema achieved? Moreover, the entanglement measure of 
superpositions of states is convex or concave as a function of the superposition 
parameter? 
Besides, can these convexity properties be predicted from the knowledge of the states 
that are being superposed?

Some of us started to contemplate these questions after working with the 
Calogero 
model \cite{Garagiola2016}. The Calogero model has many well documented 
properties and applications \cite{varias-Calogero}, but the one that caught our 
attention and was the reason to start this work is the following: the 
two-particle two dimensional case has a degenerate ground state if the total 
wave function considered is anti-symmetric under particle permutation. Most 
commonly, the eigenfunctions are chosen as eigenfunctions of the angular 
momentum operator component perpendicular to the two dimensional plane where the 
system inhabits, let us call them $\psi_{\pm}$. Other usual choice for the basis 
eigenfunctions are the combinations $(\psi_+ \pm \psi_-)/\sqrt{2}$. Denoting by 
$S(\psi)$ the von Neumann entropy of the reduced density operator obtained 
tracing out one of the particles from the two-particle  density  operator, 
$\rho=|\psi\rangle\langle\psi|$, we obtained \cite{Garagiola2016} that $S(\xi)\leq 
S(\psi_+)=S(\psi_-)$ for all $\xi=\sqrt{\alpha} \psi_+ + \sqrt{1-\alpha} \psi_-$, 
where $0\leq \alpha \leq 1$. This finding led us to analyze more general examples.

Suppose that ${L_z}$ commutes with $H$, 
where $H$ is the Hamiltonian with degenerate eigenstates labeled as 
$\left|\psi_i\right\rangle$, and ${L_z}\left|\psi_i\right\rangle 
= m_i \left|\psi_i\right\rangle$, where $m_i$ are some eigenvalues. 
Consider a superposition $\chi = \sum_i \sqrt{\alpha_i} \psi_i$, such that  $S(\chi)$ is 
convex or concave as a function of the parameters $\alpha_i$, which satisfy that 
$\sum_i \alpha_i =1$ and $0\leq \alpha_i \leq 1$.  
In this work we study only superpositions of degenerate states with eigenvalues 
of the form $\pm m \hbar$. We give a criterion that predicts the 
convexity or concavity of the entropy of the superposition. To avoid the 
uncertainty and difficulties that arise when numerical solutions are required, 
we consider a two-particle exactly solvable model, two-interacting harmonic 
oscillators; a quasi-exactly solvable one, the spherium \cite{spherium,dehesa_spherium}; a 
one-particle exactly solvable model, the Laguerre-Gaussian one photon wave 
function~\cite{kok}; and the sum of two angular momentum operators. This set of 
examples has 
been chosen because the one-party 
reduced density matrix can be exactly calculated and its eigenvalues can be 
obtained to arbitrary precision. For bipartite two-particle models one particle must 
be traced out from the whole two-particle density matrix, while for the 
one-particle model what is traced out is one of the two relevant spatial degrees 
of freedom. This procedure has been used to assess the separability of a given 
wave function as a product of two functions that depend on separate variables 
\cite{Garagiola2018} and has been also considered to assess the entanglement 
between degrees of freedom of one particle systems, for instance polarization 
and spatial 
degrees of freedom in one-photon states~\cite{one-particle-entanglement} or 
Rydberg-like harmonic states~\cite{dehesa_rydberg}.

The paper is organized as follows, in Section~\ref{sec:criterion} a criterion
to predict the convexity properties of the entropy of superposition of eigenstates is presented. 
In Sections \ref{sec:two-oscillators}, \ref{sec:spherium}, \ref{sec:lg-states} 
and 
\ref{sec:am-criterion}
we present the calculation of the eigenstates, entropies and the criterion for 
two 
interacting oscillators,
 two electrons confined in the surface of a sphere (the spherium), one photon Laguerre-Gaussian 
states 
and the addition of two angular momentum operators, respectively. We defer a 
number of mathematical details to the Appendices. Finally in 
Section~\ref{sec:discussion} we discuss our results and perspectives of the 
research.

\section{The criterion}\label{sec:criterion}

Consider a bipartite composite system with Hilbert space 
$\mathcal{H} = \mathcal{H}_A \otimes \mathcal{H}_B$, where $\mathcal{H}_A$ and $\mathcal{H}_B$ are the 
Hilbert spaces of the two subsystems $A$ and $B$, whose dimensions are equal, 
$dim(\mathcal{H}_A) = dim(\mathcal{H}_B)$.

For a given pure state in $\mathcal{H}$, $\left|\psi\right\rangle$, the 
reduced density matrices $\rho_A$ and $\rho_B$ are given by

\begin{equation}\label{eq:rho_ab}
\rho_A = \mbox{Tr}_B (\left| \psi \right\rangle \left\langle \psi \right|)   , 
\quad  \rho_B = \mbox{Tr}_A (\left| \psi \right\rangle \left\langle \psi 
\right|)  .
\end{equation} 

Both reduced density matrices are isospectral and have associated an eigenvalue problem, {\em i.e.}

\begin{equation}
\rho_A \varphi_i^A = \lambda_i^A \varphi_i^A , 
\end{equation}

\noindent and 

\begin{equation}
\rho_B \varphi_i^B = \lambda_i^B \varphi_i^B , 
\end{equation}

\noindent where $\lambda_i^A$ ($\lambda_i^B$) are the eigenvalues 
and $\varphi_i^A$ ($\varphi_i^B$)  the eigenvectors of $\rho_A$ ($\rho_B$). It 
is convenient to introduce other two  Hilbert spaces, $V_A=span(\lbrace 
\varphi_j^A \rbrace)$ where the only eigenvectors $\varphi_j^A$ that enter in 
the $span$ are those such that its corresponding eigenvalues satisfy 
$\lambda_j^A > 
0$, and equivalently for $V_B$. 

The considerations over the Hilbert spaces defined above 
become clear when computing the quantum relative entropy, a quantity commonly used to compare two quantum 
states, $\rho$ and $\sigma$. The quantum relative entropy of $\rho$ with 
respect to $\sigma$ is given by~\cite{relative-entropy}

\begin{equation}
 S(\rho \| \sigma) = -\mbox{Tr} \left( \rho \log \sigma \right) - S(\rho) = 
\mathrm{Tr} \rho (\log \rho - \log \sigma ),
\end{equation}

\noindent
where $S(\rho)$ is the von Neumann entropy

\begin{equation}
 S(\rho) = - \mbox{Tr} \rho \log \rho .
\end{equation}

If $\mbox{supp}(\rho) \bigcap \mbox{ker}(\sigma) \neq 0$ then $S(\rho\|\sigma) 
= \infty$, where $\mbox{supp}(\rho)$ y $\mbox{ker}(\sigma)$ stand for the 
support and kernel of the $\rho$ and $\sigma$ states, respectively. The 
divergence and other properties of the relative entropy can be analyzed more 
directly using the spectral decompositions

\begin{equation}\label{eq:decompositions}
 \rho = \sum_i p_i |v_i\rangle \langle v_i|, \quad \sigma = \sum_i q_i |w_i\rangle \langle w_i|,
\end{equation}

\noindent
in terms of which the quantum relative entropy can be calculated as

\begin{equation}\label{eq:relative-entropy-explicit}
 S(\rho \| \sigma) = \sum_i p_i \left( \log p_i - \sum_j (\log q_j) P_{ij} 
\right),
\end{equation}

\noindent where $P_{ij} = |\langle v_i|w_j\rangle|^2$.

Some remarks are in order. If the states $\rho$ and $\sigma$ have degenerate 
eigenvalues, each one of the decompositions in Equation~\ref{eq:decompositions} are not 
unique, since the corresponding eigenvectors $|v_i\rangle$ or $|w_i\rangle$ can be chosen in 
different ways. This possibility leads to the undesirable result that the 
relative entropy could depend on a particular election of the eigenvectors that 
corresponds to a degenerate eigenvalue. Since we intend to introduce an 
information-like quantity that
allow us to compare the reduced density matrices of different superpositions of 
Hamiltonian eigenstates, the quantity to be defined must be calculable even 
when the two drawbacks mentioned above are present.

If $H$ is the Hamiltonian of the composite system, we will focus in some 
particular 
superpositions of degenerate bound states which are also eigenstates of the $z$ 
component 
of the total angular momentum $L_z$. In particular, if 
$\left|\psi_0\right\rangle $ and $\left|\psi_1\right\rangle$ are two such eigenstates, we 
will consider the superposition 

\begin{equation}\label{eq:superposition}
\left|\psi_{\alpha} \right\rangle = \sqrt{\alpha} \left|\psi_0\right\rangle + 
\sqrt{1-\alpha} \left|\psi_1\right\rangle \quad ; \quad 0\leq \alpha \leq 1.
\end{equation}

Let us define the {\em not-shareable entropy}  of $\rho_{A,0}$ with respect to
$\rho_{A,1} $ as

\begin{equation}\label{eq:not-shareable-entropy}
S_{ns}(\rho_{A,0}) = - \sum_{\lambda_i^{A,0}>0} \Theta\left[\lambda_i^{A,0}  -  
\left\langle \rho_{A,1}\right\rangle_i \right] \log_{2}(\lambda_i^{A,0}) ,
\end{equation}

\noindent where
\begin{equation}
\rho_{A,\alpha} = \mbox{Tr}_B( \left|\psi_{\alpha} \right\rangle \left\langle 
\psi_{\alpha} \right|) ,
\end{equation}

\noindent $\left\langle \rho_{A,1}\right\rangle_i= \mbox{Tr}(P^{A,0}_i 
\rho_{A,1})$, $P^{A,0}_i$ is a one-dimensional projector associated to   
$\lambda_i^{A,0}$, and $\Theta[x]=x\,\theta(x)$, where $\theta(x)$ is the 
Heaviside 
step function. It is clear 
that when the eigenvalues of $\rho_{A,0}$ are degenerate the projectors 
$P^{A,0}_i$ are not uniquely defined. So, decomposing the sum in Equation 
\ref{eq:not-shareable-entropy} in terms of the degenerate and non-degenerate 
eigenvalues we get that
\begin{eqnarray}
 S_{ns}(\rho_{A,0}) &=& - \sum_{i} 
  \sum_{\nu=1}^{deg(\lambda_{i}^{A,0})} \Theta\left[\lambda_{i}^{A,0}  -  
  \left\langle \rho_{A,1}\right\rangle_{i,\nu} \right] 
\log_{2}(\lambda_{i}^{A,0}),  
\end{eqnarray}
where the first sum runs over the different eigenvalues and the second one over 
the degeneracy of the corresponding eigenvalue. With the previous definitions, 
we now define the entropy in which we will found our study. The {\em not-shared} 
entropy 
 of $\rho_{A,0}$ with respect to
$\rho_{A,1} $
is given by
\begin{eqnarray}\label{eq:not-shared}
S_{NS}(\rho_{A,0}) &=& - \sum_{i} 
  \mbox{min}\left\lbrace\sum_{\nu=1}^{deg(\lambda_{i}^{A,0})} 
\Theta\left[\lambda_{i}^{A,0}  -  
\left\langle \rho_{A,1}\right\rangle_{i,\nu} \right]\right\rbrace 
\log_{2}(\lambda_{i}^{A,0}), 
\end{eqnarray}

\noindent where the minimum  must be obtained in {\em each} degenerate subspace 
associated 
to a degenerate eigenvalue of $\rho_{A,0}$. The minimum can be obtained 
analytically for low degeneracy proceeding as follows. To simplify the notation 
let us call $\lambda$ the degenerate eigenvalue of interest and $\lambda_1$ 
and $\lambda_2$ the eigenvalues of $\rho_{A,1}$, it is clear that we are 
assuming 
a twofold degeneracy. In the corresponding subspace, $\left\langle 
\rho_{A,1}\right\rangle_{1} + \left\langle \rho_{A,1}\right\rangle_{2} = 
\lambda_1 + \lambda_2$, so
\begin{equation}
\tilde{S} = \min\left\lbrace\Theta\left[\lambda - \left\langle 
\rho_{A,1}\right\rangle_{1}\right] 
+  \Theta\left[\lambda - \left\langle \rho_{A,1}\right\rangle_{2}\right]\right\rbrace
\log_{2}(1/\lambda), 
\end{equation}
\noindent can be calculated explicitly as

\begin{equation}\label{eq:tilde-entropy}
\tilde{S} = \left\lbrace 
\begin{array}{lcl}
(2 \lambda - (\lambda_1 + \lambda_2)) \log_{2}(1/\lambda)& \qquad&  \mbox{if} 
\quad \lambda > \displaystyle{\frac{\lambda_1 + \lambda_2}{2}} \\
0 && \mbox{otherwise}
\end{array} \right. .
\end{equation}

Once the not-shared entropy is defined we introduce, in a similar fashion, the 
{\em remaining} entropy of state $\rho_{A,0}$, which is given by

\begin{equation}\label{eq:remaining-entropy}
S_{R}(\rho_{A,0}) = S(\rho_{A,0}) - S_{NS}(\rho_{A,0}) .
\end{equation}

We are now in conditions to state the criterion about the convexity properties 
of 
the entropy of a superposition of two degenerate states, like the one defined in 
Equation \ref{eq:superposition}. If both states, $\psi_0$ and $\psi_1$, are eigenfunctions of the $z$ 
component of the total angular momentum operator, $L_z$, 
with eigenvalues $\pm m\hbar$, then

\begin{equation}\label{eq:criterion1}
S(\rho_{A,\alpha}) \leq \alpha\, S(\rho_{A,0})+ (1-\alpha)S(\rho_{A,1}) 
\quad \mbox{if} \quad S_{NS}(\rho_{A,0}) < S_R (\rho_{A,0})
\end{equation}

\noindent and

\begin{equation}\label{eq:criterion2}
S(\rho_{A,\alpha}) \geq \alpha\, S(\rho_{A,0}) + (1-\alpha) S(\rho_{A,1}) 
\quad \mbox{if} \quad S_{NS}(\rho_{A,0}) > S_{R} (\rho_{A,0}) .
\end{equation}

\noindent  The criterion predicts exactly the convexity properties expected for the entropy, and 
this can be noted by the quantity,
\begin{equation}\label{eq:crit}
Q_c=sgn(S_{R}(\rho_{A,0})-S_{NS}(\rho_{A,0})),
\end{equation}

\noindent which has the value $+1$ when the entropy $S(\rho_{A,\alpha})$ is convex and $-1$ when it is concave.

Then, the criterion establishes a relationship between the convexity of the entropy of a superposition state with 
an entropic quantity that depends only on the
extremal states of such superposition. In the following sections we will test 
the 
criterion in several systems
and show how it correctly predicts how state superposition can lead to more or less 
information entropy.

It is worth to point that the criterion can be stated also in 
terms of the state $\rho_{A,1}$ with respect to the state $\rho_{A,0}$. 
Even more, its seems desirable to state a criterion in which both states play 
more symmetrical roles. For the moment we prefer to use the criterion as stated 
from Equation~\ref{eq:not-shareable-entropy} trough  Equation~\ref{eq:crit}. The 
criterion compares 
a given state, say $\rho_{A,0} $ with another, say $\rho_{A,1} $, so the {\em 
not-shared} entropy contains information of $\rho_{A,0} $ that can not be 
obtained from state $\rho_{A,1} $ measuring the last in the basis of 
eigenprojectors of the former. If $V_A\bigcap V_B = \phi$  then 
$S_{NS}(\rho_{A,0}) = S(\rho_{A,0})$. In this sense, the not-shared  entropy 
measures how different are the states when both are measured using the 
projectors associated to one of them.

The \emph{remaining} entropy quantifies how much information can be obtained 
about $\rho_{A,0}$ 
when $\rho_{A,1}$ is measured using the projectors associated to $\rho_{A,0}$. 
This can be seen rather directly when both states, $\rho_{A,0} $ and 
$\rho_{A,1}$, have non-degenerate eigenvalues and the same set of eigenvectors. 
In that case

\begin{equation}\label{eq:simple-remaining}
S_R = -\sum_i \min{\left(\lambda_i^{A,0}, \lambda_i^{A,1}\right)} \log_2(\lambda_i^{A,0}),
\end{equation}
and the sum runs over the eigenvalues $\lambda_i^{A,0}$ whose eigenvectors 
belong to $V_A\bigcap V_B $.

{ The not-shared entropy, as proposed in 
Equation~\ref{eq:not-shared}, depends on the spectral decomposition of both 
reduced density operators $\rho_{A,0}$ and $\rho_{A,1}$. This dependency does 
not imply a restriction from the point of view of the calculations in concrete 
systems but, as is the case for entanglement measures, a definition in terms 
of a minimization over a Hilbert space can offer a more general point of view or 
an operational procedure to determine the not-shared entropy. We will return to 
this subject once the example considering states in finite dimensional Hilbert 
spaces, Section~\ref{sec:am-criterion}, has been analyzed thoroughly.} 

In the next Sections we apply the criterion to several bipartite systems. The 
different 
Hamiltonians allow us to carry the calculations to obtain the reduced density 
matrices analytically.

\section{Two interacting two-dimensional oscillators}\label{sec:two-oscillators}

A well known exactly solvable problem consists of two particles interacting harmonically confined in 
an harmonic trap.
 We use units such that $\hbar=1$, $m=1$, $\omega=1$, where $m$ is the mass and $\omega$ is the 
 frequency of both oscillators. The Hamiltonian is given by
 
\begin{equation}\label{eq:two-dimensional-oscillators}
H(\vec{x_1},\vec{x}_2)=-\frac{1}{2}\bigtriangledown_1^2 -\frac{1}{2} 
\bigtriangledown_2^2 +\frac{1}{2}|\vec{x}_1|^2+\frac{1}{2}|\vec{x}_2|^2+\lambda|\vec{x}_1-\vec{x}_2 |^2
\end{equation}

\noindent where $\vec{x}_i=(x_i,y_i)$,  $\bigtriangledown_i^2$ is the 
two-dimensional
Laplacian, and $i=1,2$ is the index numbering the particles.  

The parameter $\lambda$ allows to switch from a non-interacting system, 
$\lambda=0$, 
to an interacting one, $\lambda>0$. 
The symmetries and quantum numbers of the wave functions can be chosen at convenience. Since the superpositions in which we are interested are composed 
of 
eigenfunctions 
whose energy eigenvalues are degenerate but with different angular momentum eigenvalue,
it is useful to derive expressions for the wave functions in 
coordinates where the system can be recast as a  
non-interacting one. So, introducing the centered and relative coordinates 

\begin{equation}
\vec{R}=\vec{x}_1+\vec{x}_2, \qquad \vec{r}=\vec{x}_1-\vec{x}_2,
\end{equation}
respectively, the Hamiltonian in Equation~\ref{eq:two-dimensional-oscillators} 
can 
be written as

\begin{eqnarray}
\label{eq:two-oscillators-relative}
H(\vec{R},\vec{r}) &=& H_{\omega_R}(\vec{R}) + H_{\omega_r}(\vec{r}) \\ \nonumber
&=& -\bigtriangledown_R^{\,2}+\frac{1}{4}
|\vec{R}|^2- 
\bigtriangledown_r^2+(\frac{1}{4}+\lambda)|\vec{r}|^{\,2} ,
\end{eqnarray}

\noindent where $H_{\omega}$ is a two-dimensional harmonic oscillator Hamiltonian, 
with frequency $\omega$ and mass equal $1/2$. The oscillator that 
depends on the relative coordinates has $\omega_r=\sqrt{4\lambda+1}$, while 
the other, that depends on the centered coordinates, has $\omega_R=1$. The exact eigenfunctions and eigenvalues 
of the one-particle Hamiltonians in Equation \ref{eq:two-oscillators-relative} are well known in several 
coordinate systems~\cite{Cohen}. The two-particle eigenfunctions can now 
be written as a product of a pair of one-particle ones.

The one-particle cylindrical eigenvectors, $\left| m,n\right\rangle$, satisfy

\begin{eqnarray}
H_{\omega}\left| n,m \right\rangle &=& (2n+|m|+1)\omega \left| n,m \right\rangle,\\ \nonumber 
L^{(1)}_z \left| n,m \right\rangle &=& m \left| n,m \right\rangle,
\end{eqnarray}

\noindent where $\omega$ is either $\omega_R$ or $\omega_r$, and $L^{(1)}_z$ is 
the $z$-component of the one-particle angular momentum. 

\begin{widetext}
Collecting the results given above, the two-particle eigenvector, 
$\left|n,m,l,p\right\rangle=\left|n,m\right\rangle\left|l,p\right\rangle$, satisfy

\begin{eqnarray}\label{eq:two-oscillators-eigenvalues}
H\left|n,m,l,p\right\rangle &=& 
\left[(2n+|m|+1)\omega_R+(2l+|p|+1)\omega_r\right]
\left|n,m,l,p\right\rangle,
\end{eqnarray}

\noindent and

\begin{equation}\label{eq:two-oscillators-angular-momentum}
L_z \left|n,m,l,p\right\rangle = (m+p) \left|n,m,l,p\right\rangle ,
\end{equation}
\end{widetext}

\noindent where $L_z=L_{z_R}+ L_{z_r}$.

Equations~\ref{eq:two-oscillators-eigenvalues} and 
\ref{eq:two-oscillators-angular-momentum} show that, when $\sqrt{4\lambda+1}$ is a natural number, there are many different ways to combine the quantum numbers 
$n,m,l$ and $p$ to obtain degenerate eigenfunctions with different values of 
angular momentum quantum number. 

The traces in Equation~\ref{eq:rho_ab} over $A$ and $B$ have the meaning of
particle $1$ and $2$. Hence the reduced density matrix of a particular
angular momentum state is

\begin{equation}\label{eq:rho_nmlp}
\rho_A(\vec{x}_1,\vec{x}_1^{\prime}) = \int d^2x_2\,  \psi^*_{n,m,l,p}(\vec{x}_1,\vec{x}_2)\psi_{n,m,l,p}(\vec{x}_1^{\prime},\vec{x}_2).
\end{equation}

\noindent The states $\left|\psi_0\right\rangle$ and $\left|\psi_1\right\rangle$ 
are 
now identified with two different angular momentum states 
$\left|n,m,l,p\right\rangle$ that have the same energy and opposite eigenvalues for $L_z$. 
The reduced density matrix for the mixed state $\rho_{A,\alpha}$ is defined 
analogously to $\rho_A(\vec{x}_1,\vec{x}_1^{\prime})$ in 
Equation~\ref{eq:rho_nmlp}. We explain how to obtain an exact expression
for $\rho_A(\vec{x}_1,\vec{x}_1^{\prime})$  and how to compute its eigenvalues 
in Appendix \ref{ap:two-oscillators}.

\subsection{Non-interacting harmonic oscillators, $\lambda=0$}

\begin{figure}
    \centering
    \includegraphics[height=6cm]{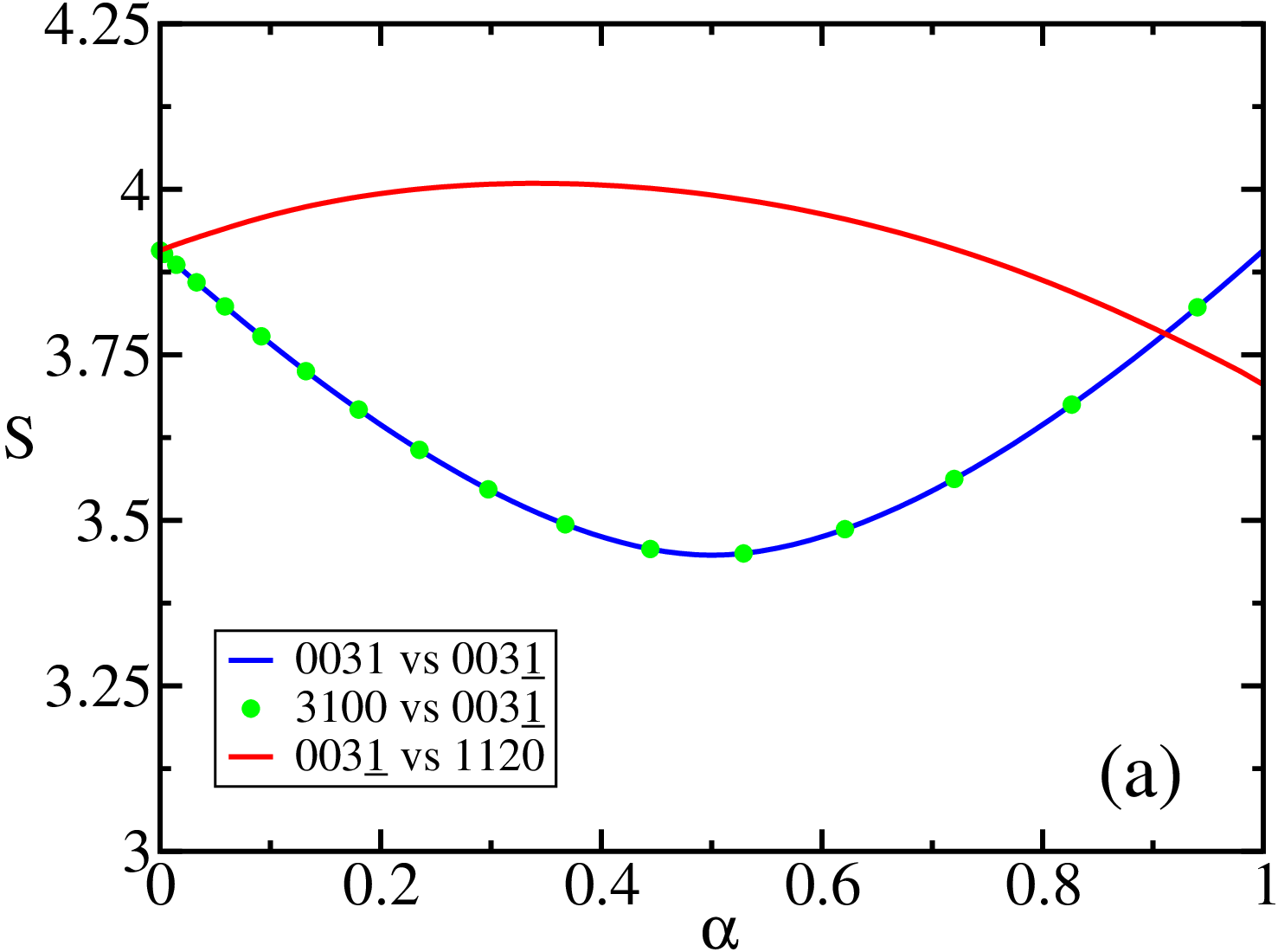}
    \includegraphics[height=6cm]{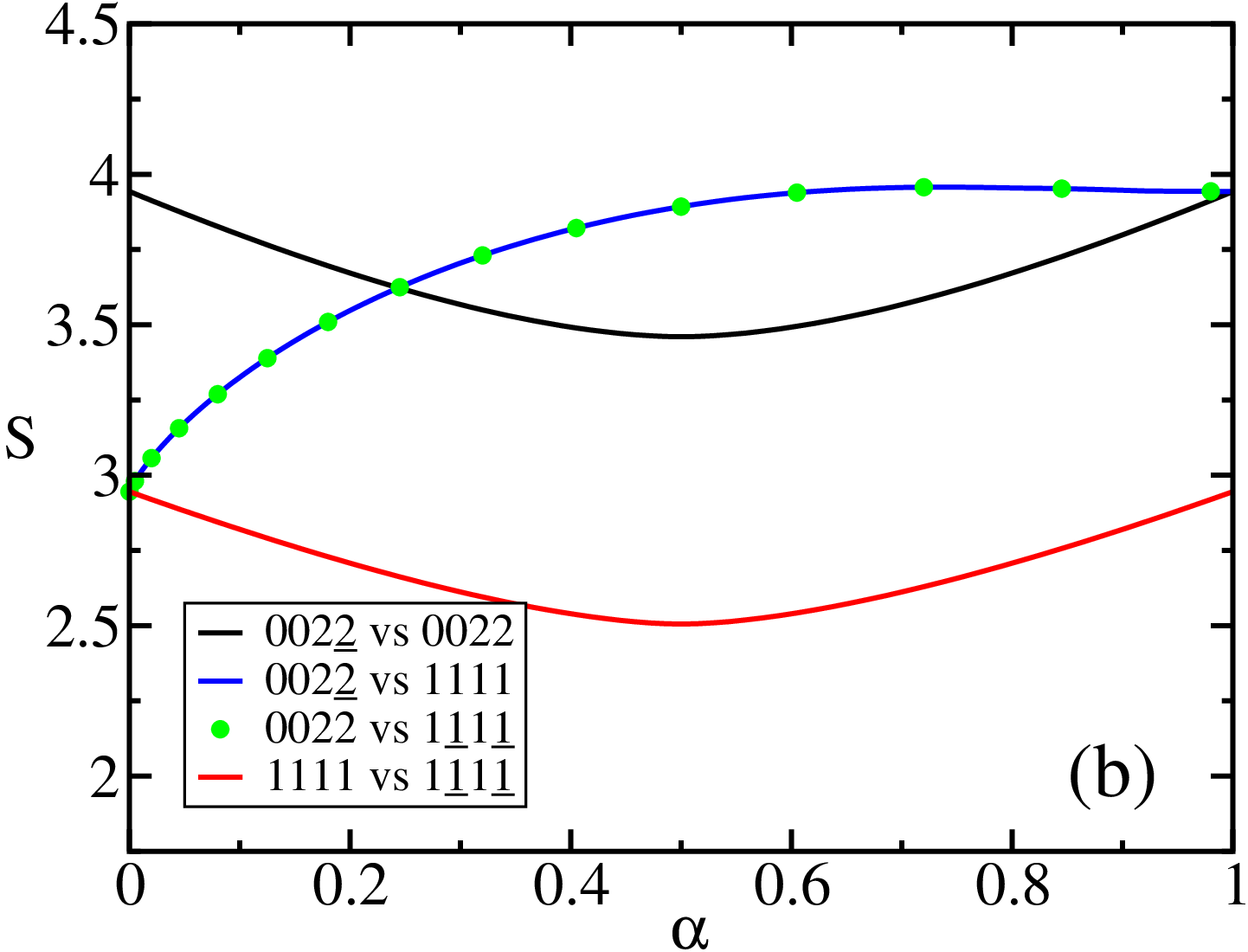}
    \caption{\label{fig:non-interacting} von Neumann entropy as a function of the 
    superposition parameter $\alpha$, for states $\left|\psi_{\alpha}\right\rangle =\sqrt{\alpha}\left|n,m,l,p\right\rangle 
+ \sqrt{1-\alpha}\left|n',m',l',p'\right\rangle$ with $\lambda=0$ 
(non-interacting oscillators). Note that a negative quantum number, $-n$, is 
denoted as $\underline{n}$ in the legends. (a) For $|m+p|=1$,  the 
superpositions of states that are shown are $\left|1,1,2,0\right\rangle$ with 
$\left|0,0,3,-1\right\rangle$ (red full),
$\left|0,0,3,-1\right\rangle$ with $\left|0,0,3,1\right\rangle$ (blue full), 
$\left|0,0,3,-2\right\rangle$ with $\left|3,1,0,0\right\rangle$ (green dots). 
(b) For $|m+p|=2$, the superpositions of states that are shown correspond to 
$\left|0,0,2,-2\right\rangle$ with 
$\left|0,0,2,2\right\rangle$ (black full), $\left|0,0,2,-2\right\rangle$ with 
$\left|1,1,1,1\right\rangle$ (blue full),
$\left|0,0,2,2\right\rangle$ with $\left|1,-1,1,-1\right\rangle$ (green dots) 
and $\left|1,1,1,1\right\rangle$ with $\left|1,-1,1,-1\right\rangle$ (full red). 
See the corresponding values of $Q_c$ in Table~\ref{ta:non-interacting}.}
\end{figure}

\begin{table}[h!]
\centering
\begin{tabular}{|c|c|c|c|c|c|c|}
\hline 
$nmlp$ & $n'm' l' p'$ & convexity & $Q_c$ &$S_{vn}$ &  $S_{NS}$ &$S_{R}$  \\ 
\hline 
\multicolumn{7}{|c|}{$|m+p|=1$}  \\
\hline
003\n{1} & 0031 & convex & + & 3.907 &  0 &3.907  \\ 
\hline 
003\n{1} & 3100 & convex & + &3.907 &  0 &3.907 \\ 
\hline 
1120 & 003\n{1} & concave & -- & 3.704 & 1.959 & 1.745  \\ 
\hline 
\multicolumn{7}{|c|}{$|m+p|=2$}  \\
\hline
002\n{2} & 0022 & convex & + &3.94 & 0 & 3.94\\
\hline
002\n{2} & 1111 & concave & -- & 3.94 & 2.85 & 1.09\\
\hline
0022 & 1\n{1}1\n{1} & concave & -- &3.94 & 2.85 & 1.09\\
\hline
1111 & 1\n{1}1\n{1} & convex & + & 2.94 & 0 & 2.94\\
\hline
\end{tabular} \\
\caption{\label{ta:non-interacting} Entropies, convexity and the $Q_c$ value  
for the two cases of non-interacting oscillators, $\lambda=0$, shown in 
Figure~\ref{fig:non-interacting} ($|m+p|=1$ and $|m+p|=2$). Note that in the first two columns a negative quantum number, $-n$, is denoted as $\underline{n}$. }
\end{table}

The bipartite states are labeled with the four quantum numbers $n,m,l,p$. In this section we 
will 
consider superpositions of states that satisfy $|\,m+p\,|=1$ or $|\,m+p\,| =2$, 
there is no \emph{a priori} restriction to the values
of the quantum numbers $n$ and $l$ but the one imposed by 
Equation~\ref{eq:two-oscillators-eigenvalues}.
However, since we use a finite basis from which we obtain the eigenvalues 
of $\rho_{A,\alpha}$, we consider the states for $n,l\leq 3$.



Figure~\ref{fig:non-interacting} shows the behaviour of the von Neumann entropy 
as 
a function of the superposition parameter $\alpha$. Both possible curvatures, 
or convexities, can be clearly appreciable according with the states that enter 
in the superposition. In Table~\ref{ta:non-interacting} the values of von 
Neumann,
remaining and not-shared entropies, and $Q_c$ are listed for all the cases 
shown in Figure 1. 
The 
criterion predicts correctly what convexity is to be expected.

\subsection{Interacting harmonic oscillators, $\lambda\neq0$}

Despite that a two-particle system, both confined by an harmonic potential and 
interacting harmonically, is exactly solvable, the evaluation of the necessary 
matrix representation of a given wave function (see Appendix 
\ref{ap:two-oscillators}) becomes quite taxing. Because of this, it is simpler, 
and computationally faster, to obtain very accurate approximate variational wave functions 
using basis set functions that are products of one-particle functions. Once 
these variational eigenfunctions are calculated they can be used to evaluate 
matrix representations and approximate eigenvalues of the reduced density 
matrix corresponding to a given variational eigenfunction.  The procedure to 
calculate each one of these quantities has been described elsewhere, see for 
instance~\cite{Garagiola2016,Ho} and References therein. The accuracy of 
the whole procedure can be assessed in different ways, mainly comparing the 
variational eigenvalues with the corresponding exact values computed using 
Equation~\ref{eq:two-oscillators-eigenvalues}. For all the cases that are 
discussed in this Section the variational eigenvalues corresponding to the 
variational eigenfunctions used to construct superpositions differ from the 
exact ones in less than $1\times 10^{-4}$.

\begin{figure}[h]
\centering
\includegraphics[height=7cm]{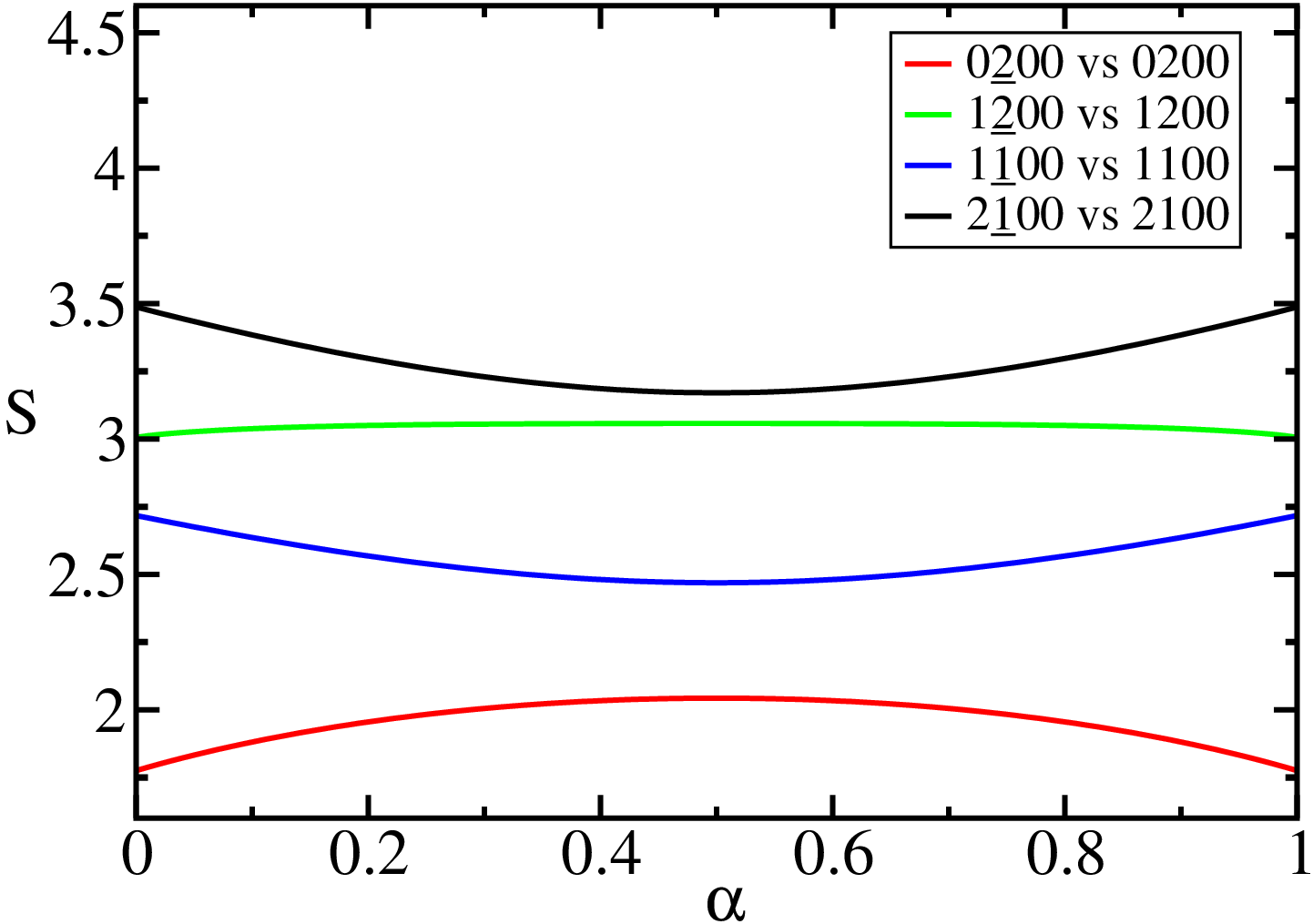}
\caption{\label{fig:interacting-oscillators}von Neumann entropy as a function 
of 
the combination
parameter $\alpha$ for states $\left|\psi_{\alpha}\right\rangle 
=\sqrt{\alpha}\left|n,m,l,p\right\rangle 
+ \sqrt{1-\alpha}\left|n',m',l',p'\right\rangle$ of two interacting 
oscillators with 
$\lambda=0.7$ and $|m+p|=1,2$. The combinations of states that are shown are 
$\left|0,-2,0,0\right\rangle$ with $\left|0,2,0,0\right\rangle$ 
(red), $\left|1,-2,0,0\right\rangle$ with $\left|1,2,0,0\right\rangle$ 
(green), $\left|1,-1,0,0\right\rangle$ with $\left|1,1,0,0\right\rangle$ 
(blue), and $\left|2,-1,0,0\right\rangle$ with $\left|2,1,0,0\right\rangle$ 
(black).}
\end{figure}

\begin{table}[h!]
\centering
\begin{tabular}{|c|c|c|c|c|c|c|}
\hline 
$nmlp$ & $n'm' l' p'$ & convexity & $Q_c$ &$S_{vn}$ &  $S_{NS}$ &$S_{R}$  \\ 
\hline 
\multicolumn{7}{|c|}{$|m+p|=1$}  \\
\hline
1\n{1}00 & 1100 & convex & + & 2.717 & 1.034  & 1.683 \\ 
\hline 
2\n{1}00 & 2100 & convex & + & 3.487&  1.121 & 2.366\\ 
\hline
\multicolumn{7}{|c|}{$|m+p|=2$}  \\
\hline

0\n{2}00 & 0200 & concave & -- & 1.776& 1.123 & 0.653\\
\hline
1\n{2}00 & 1200 & concave & -- & 3.006& 1.740 & 1.266\\
\hline
\end{tabular} \\
\caption{\label{ta:interacting} Entropies, convexity and the $Q_c$ value  
for the two cases of interacting oscillators, $\lambda=0.7$, shown in 
Figure~\ref{fig:interacting-oscillators} ($|m+p|=1$ and $|m+p|=2$).}
\end{table}

One of the  drawbacks of the variational method comes from the fact that 
assigning quantum numbers to the variational eigenfunctions is usually 
complicated. In the present case, this assignment is simplified by choosing a 
particular value of the interaction parameter, $\lambda=0.7$, that separates 
adequately the frequencies $\omega_R=1$ and $\omega_r\simeq 1.949$, and using 
basis sets with well defined values of $L_z$. All in all, for the lowest 
eigenvalues it is possible to unequivocally label the variational eigenfunction 
using the set of quantum numbers  in 
Equation~\ref{eq:two-oscillators-eigenvalues}. The results 
for the interacting system can then be presented using the same conventions that 
those used for  the non-interacting one. The von Neumann entropy as a 
function of the superposition parameter is shown in 
Figure~\ref{fig:interacting-oscillators} for several superpositions 
and the corresponding values for the remaining entropy,  the not-shared one 
and the criterion are collected in Table~\ref{ta:interacting}.

The current example shows all the features that are characteristic of what can
be observed in superpositions of degenerate states. The changes in the 
convexity 
are correctly predicted by the criterion presented in 
Section~\ref{sec:criterion}. 
The example is valuable because it is exactly soluble and its numerical (variational) implementation is simple.  But, to some extent, 
the peculiar behaviour of systems of harmonic oscillators, 
that can be cast as interacting or non-interacting, 
limits the scope and validity of the example. So in Sections~\ref{sec:spherium}
and \ref{sec:lg-states} we present further examples that can be analyzed 
analytically and present, in the case of the Spherium model,  strong 
correlations between the particles.

\vspace{3mm}

\section{The Spherium model}\label{sec:spherium}

The number of exactly solvable two interacting particle models in different 
spatial dimensions is remarkably low, which explains why so many studies of 
entanglement entropies are about systems of harmonic oscillators or variants of 
the Calogero model. As has been said in the Introduction, the study of 
the Calogero model was what triggered the formulation of the criterion that is 
the object of the present work. Fortunately, there is a growing number of 
quasi-exactly solvable models \cite{quasi-exactly-varios} that can be used to 
study properties of strongly interacting two-particle models \cite{Pont2018}. 

The spherium model~\cite{Spherium-varios}, \emph{i.e.} two electrons interacting via the Coulomb 
potential and confined to the surface of a $D$ dimensional sphere, was 
proposed to 
study the properties of electronic correlations in a confining geometry. It has 
been studied using different approaches and as a benchmark to test numerical 
approximations. 
As was shown in Reference~\cite{spherium}, this model is quasi-exactly solvable, with analytical solutions
for particular values of the radius $R$ of the sphere and the dimension $D$. 
These solutions can be found writing the two-electron wave function as a bipolar 
expansion~\cite{bipolar}, which allows to calculate eigenfunctions with small total 
orbital angular momentum numbers, $L=0,1$ or $2$, but it is quite cumbersome to 
implement for larger values of $L$. 

To test the criterion we will construct two-electron wave functions with $L=0,1$ and $2$, 
following the work of Pestka~\cite{Petska}, which can be applied in a systematic way. 
In this Section we present the main details of the derivation of the wave 
functions, the reduced matrix elements and its eigenvalues.

\subsection{Description of the General Solution for $D=3$}

\begin{widetext}

Consider a two-particle Hamiltonian of the form

\begin{equation}
\label{eq:peska-hamiltonian}
H=-\frac{1}{2}\nabla^2_1-\frac{1}{2}\nabla^2_2+V(r_1,r_2,r_{12}).
\end{equation}

\noindent Note that the potential 
depends on the radial coordinates of both particles, $r_1$ and $r_2$, and the 
distance between them $r_{12}$. The bipolar decomposition assumes that the 
solutions that are simultaneous eigenfunctions of the Hamiltonian,  the total 
orbital angular momentum and the $z$ component of the orbital angular momentum, 
with eigenvalues $E$, $L(L+1)$ and $M$, respectively, can be written as

\begin{equation}
\label{eq:wave-function-bi}
\psi(\textbf{r}_1,\textbf{r}_2)=\sum_{l_1=d_0}^{L}\Phi_{l_1\,l_2}(r_1,r_2,r_{12}
)\Omega_{l_1,l_2}^{L,M}(\hat{r}_1,\hat{r}_2),
\end{equation}

\noindent with the constraint $l_2=d-l_1$. The inferior limit of the sum must be determined using the parity of the solution 
$\pi=(-1)^{l_1+l_2}$, according to

\begin{equation}
d_0 = L+1-d, \qquad \mbox{and} \quad d= \left\lbrace
\begin{array}{lcl}
L+1  & \mbox{if} & \pi = (-1)^L \\
L    & \mbox{if} & \pi = (-1)^{L+1}
\end{array}
\right. .
\end{equation}

\noindent The functions $\Omega_{l_1,l_2}^{L,M}(\hat{r}_1,\hat{r}_2)$ are the eigenfunctions of $\mathbf{L}^2$ and
$L_z$

\begin{equation}
\mathbf{L}^2 \Omega_{l_1,l_2}^{L,M}(\hat{r}_1,\hat{r}_2) = L(L+1) 
\Omega_{l_1,l_2}^{L,M}(\hat{r}_1,\hat{r}_2),
\end{equation}

\noindent and 

\begin{equation}
L_z \Omega_{l_1,l_2}^{L,M}(\hat{r}_1,\hat{r}_2) = M \Omega_{l_1,l_2}^{L,M}(\hat{r}_1,\hat{r}_2).
\end{equation}

Note that the wave function in Equation~\ref{eq:wave-function-bi} has $L+1$ 
or $L$ radial functions (depending on the parity) $\Phi_{l_1\, l_2}(r_1,r_2,r_{12})$. Some 
algebra can be simplified noting that 

\begin{equation}
\label{eq:lap}
\nabla^2_i 
f(r_1,r_2,r_{12})\Omega^{L,M}_{l_1,l_2}(\hat{r}_1,\hat{r}_2) 
=\sum_{\tilde{l}_1=d_0}^{L}\{\hat{X}_i^{\tilde{l}_i,l_i}f(r_1,r_2,r_{12})\}
\Omega^ { L , M } _ { \tilde{l}_1,\tilde{l}_2}(\hat{r}_1,\hat{r}_2),
\end{equation}

\noindent where $\nabla^2_i$ is any of the Laplacian operators that enter in 
Equation~\ref{eq:peska-hamiltonian}, and $f(r_1,r_2,r_{12})$ is any function 
that depends only on the variables $r_1,r_2$ and $r_{12}$. The curly brackets 
indicate that the operators $\hat{X}_i^{\tilde{l}_i,l_i}$ are applied only over 
the radial terms.

The operators $\hat{X}_i^{\tilde{l}_i,l_i}$ have the property that

\begin{equation}
 \hat{X}_2^{\tilde{l},l}(r_1,r_2,r_{12})=\hat{X}_1^{\tilde{l},l}(r_2,r_1,r_{12}),
\end{equation}

\noindent and can be written as

\begin{eqnarray}
 \hat{X}_i^{\tilde{l}_i,l_i}&=&\frac{\partial^2}{\partial 
r_i^2}+\frac{2}{r_i}\frac{\partial}{\partial r_i} +\frac{\partial ^2}{\partial 
r_{12}^2} \\
&&+\frac{(2+l_i)r_i^2+l_i(r_{j}^2-r_{12}^2)}{r_i^2r_{12}}\frac{\partial}{
\partial r_{12}}
 +\frac{r_i^2-r_{j}^2+r_{12}^2}{r_ir_{12}}\frac{\partial ^2}{\partial r_i 
\partial r_{12}}-\frac{l_i(l_i+1)}{r_i^2}, \nonumber
\end{eqnarray}

\noindent when $0\leq l_i \leq L$ and 

\begin{equation}
\hat{X}_i^{\tilde{l}_i-1,l_i}=\frac{-2r_{j}}{r_ir_{12}}\sqrt{\frac{(L-l_{
j})(2l_i+1)(L-l_i+1)}{2l_{j}+3}}\frac{\partial}{\partial r_{12}},
\end{equation}

\noindent for $1\leq l_i \leq L$. In both cases the value of $j$ is such that $j\neq i$.

When the particles are confined to the surface of a sphere, $r_1=r_2=R$, the last 
two expressions
can be further simplified, for $0\leq l_i \leq L$

\begin{equation}
 \label{x1}
\hat{X}_i^{\tilde{l}_i,l_i}=\frac{\partial ^2}{\partial r_{12}^2} 
+\frac{(2+l_i)R^2+l_i(R^2-r_{12}^2)}{R^2r_{12}}\frac{\partial}{
\partial r_{12}},
\end{equation}

\noindent where a constant term is omitted, and for $1\leq l_i \leq L$  

\begin{equation}
\hat{X}_i^{\tilde{l}_i-1,l_i}=\frac{-2}{r_{12}}\sqrt{\frac{(L-l_{
j})(2l_i+1)(L-l_i+1)}{2l_{j}+3}}\frac{\partial}{\partial r_{12}}.
\end{equation}






All in all, the method proposed by Petska reduce the calculation of the 
two-electron 
wave function to another problem which consists in a set of $L$ (or $L+1$) 
coupled equations for the quantities $\Phi_{l_1\, l_2}$. 

\subsection{Wave functions with $L=2$}

The wave function with angular momentum number $L=2$ and parity $\pi=(-1)^{L+1}$, depends on just two radial 
functions, since there are two combinations of possible values $l_1=1$, $l_2=2$ 
and $l_1=2$, $l_2=1$. Using this, we get 

\begin{eqnarray}
 & & \Big[-\frac{1}{2}\Big(\frac{\partial ^2}{\partial 
r_{12}^2}+\frac{4R^2-r_{12}^2}{R^2r_{12}}\frac{\partial}{\partial 
r_{12}}\Big)-\frac{1}{2}\Big(\frac{\partial ^2}{\partial 
r_{12}^2}+\frac{6R^2-2r_{12}^2}{R^2r_{12}}\frac{\partial}{\partial 
r_{12}}\Big)+\frac{1}{r_{12}}-E\Big]\Phi_{12}\nonumber \\
& &+\frac{1}{r_{12}}\frac{
\partial } { \partial r_{12}}\Phi_{21} =0, \label{eq:phi12}
\end{eqnarray}

\noindent and

\begin{eqnarray}
 & & \Big[-\frac{1}{2}\Big(\frac{\partial ^2}{\partial 
r_{12}^2}+\frac{4R^2-r_{12}^2}{R^2r_{12}}\frac{\partial}{\partial 
r_{12}}\Big)-\frac{1}{2}\Big(\frac{\partial ^2}{\partial 
r_{12}^2}+\frac{6R^2-2r_{12}^2}{R^2r_{12}}\frac{\partial}{\partial 
r_{12}}\Big)+\frac{1}{r_{12}}-E\Big]\Phi_{21}\nonumber \\
&&+\frac{1}{r_{12}}\frac{
\partial } { \partial r_{12}}\Phi_{12}=0.\label{eq:phi21}
\end{eqnarray}

\noindent These pair of coupled equations have a solution of the form $\Phi_{12}=\Phi_{21}=\Phi$. In this case
Equations~\ref{eq:phi12} and \ref{eq:phi21} reduce to

\begin{equation}
 \Bigg[\frac{\partial ^2}{\partial 
r_{12}^2}+\Bigg(\frac{4}{r_{12}}-\frac{3}{2R^2}r_{12}\Bigg)\frac{\partial}{
\partial r_{12}}-\frac{1}{r_{12}}+E\Bigg]\Phi=0.
\end{equation}

Using the ansatz

\begin{equation}\label{eq:l-two-solution}
 \Phi=1+r_{12}/\alpha,
\end{equation}

\noindent we found that this function is a solution with $\alpha=4$ and $R^2=6$.

Collecting again the partial results, we get the whole set of wave functions 
with $L=2$ and $M=0,\pm 1, \pm 2$

\begin{equation}
\Psi_{2,M}(\Omega_1,\Omega_2,r_{12})=(\mathcal{Y}_{1,2}^{2,M}(\Omega_1,
\Omega_2)-\mathcal{Y}_{1,2}^{2,M}(\Omega_2,\Omega_1))(1+r_{12}/4),
\end{equation}

\noindent where $\mathcal{Y}_{1,2}^{2,M}(\Omega_1,\Omega_2)$ are the usual total angular
momentum eigenfunctions defined in terms of the spherical harmonics and the 
Clebsch-Gordan coefficients.

\end{widetext}

The one-electron reduced density matrix can be straightforwardly  obtained using
Perkins' formula~\cite{perkins}

\begin{equation}\label{eq:suma-harmonicos-esfericos}
r_{12}^k=4\pi\sum_{l=0}^{L_1^k}\Big(\sum_{m=-l}^lY_{lm}^*(\Omega_1)Y_{lm}
(\Omega_2)\Big)\Big(\sum_{t=0}^{L_2^{k,l}}C_{klt}\, r_{<}^{l+2t}\, r_{>}^{k-(l+2t)}\Big) .
\end{equation}

This expression is valid for $k=-1,0,1,..$, and the coefficients $L_1^k$ and 
$L_2^{k,l}$ 
are given by

\begin{equation}
L_1^k = \left\lbrace 
\begin{array}{cll}
  \frac{k}{2}   &  \mbox{if} & k \; \mbox{is even}\\
   \infty  & \mbox{if} & k \; \mbox{is odd}
\end{array}
\right.,
\end{equation}

\noindent and

\begin{equation}
L_2^{k,l} = \left\lbrace 
\begin{array}{cll}
  \frac{k}{2} -l  &  \mbox{if} & k \; \mbox{is even}\\
   \frac{k+1}{2}  & \mbox{if} & k \; \mbox{is odd}
\end{array}
\right. .
\end{equation}

\begin{figure}[h] 
\begin{center}
\includegraphics[height=7cm]{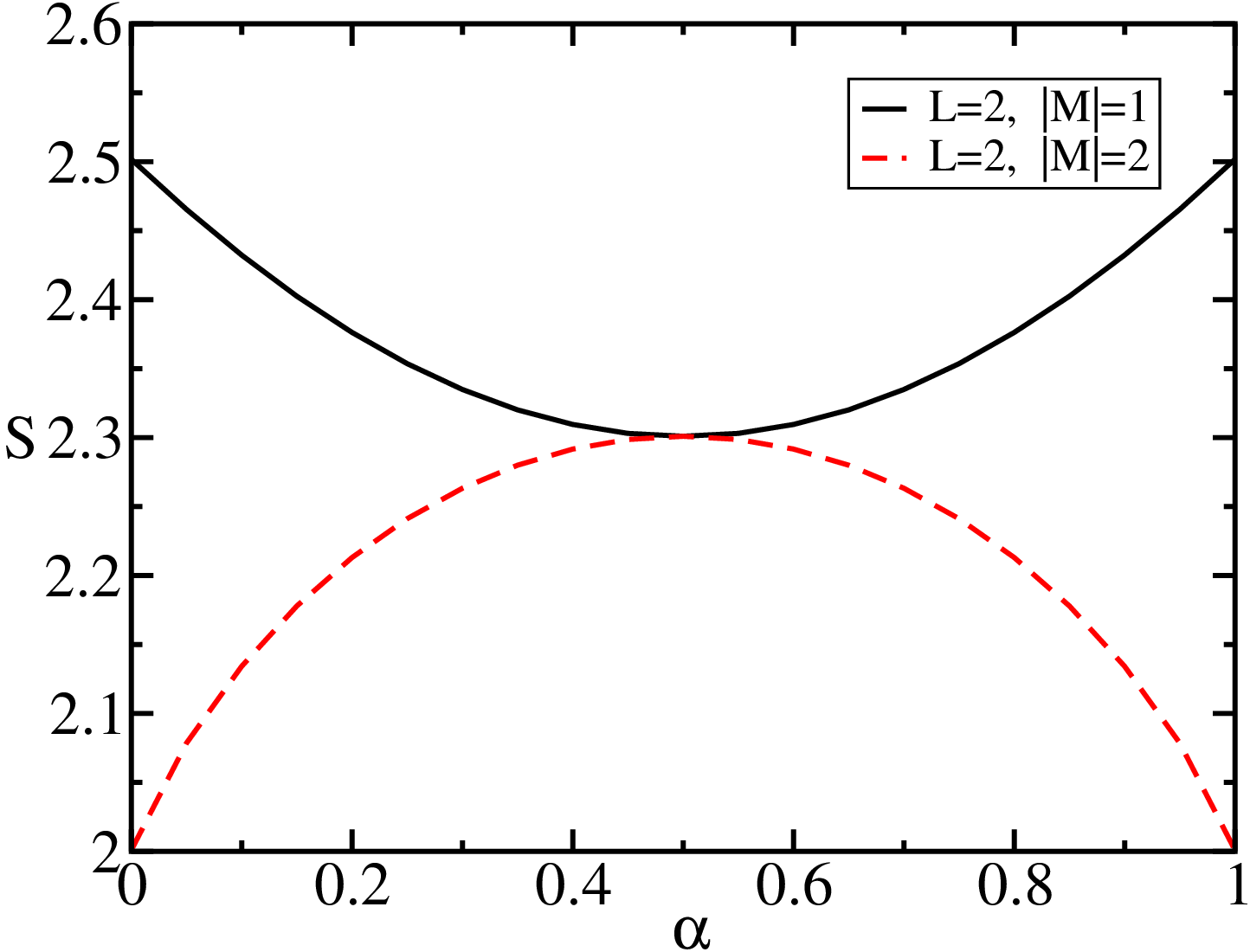} \\ 
\end{center}
\caption{\label{fig:petska} The von Neumann entropy as a function of $\alpha$
for $L=2$ states of two electrons on a surface of a sphere. The states are a linear
combination of the form shown in Equation~\ref{eq:lc_petska}. The entropy with
$|M|=1$ is depicted with the black line and that with $|M|=2$ with a red dashed line.
In Table \ref{ta:petska} the values of the entropies, the criterion prediction 
and 
convexities of the curves
are shown.}
\end{figure}

\begin{table}[h]
\centering
\begin{tabular}{|c|c|c|c|c|c|}
\hline 
$LM$ & $LM'$ & convexity & $Q_c$ &$S_{NS}$ & $S_{R}$ \\ 
\hline 
21 & 2\underline{1} & convex & + & 0.917 &1.584 \\ 
\hline
22 & 2\underline{2} & concave & -- &1.422 &0.578 \\ 
\hline
\end{tabular}
\caption{\label{ta:petska} The values of the convexity, the criterion 
prediction 
and entropies of the curves
shown in Figure~\ref{fig:petska}.}
\end{table}

The coefficients $C_{klt}$ are given by
\begin{equation}
 C_{klt}=\left\lbrace
 \begin{array}{lr}
 \displaystyle{\frac{1}{k+2}{{k+2}\choose{2t+1}}} & \quad \mbox{if}\quad l=0 \\
  \displaystyle{\frac{1}{k+2}{{k+2}\choose{2t+1}} \prod_{\alpha=0}^{\min[l-1,(k+1)/2]}\frac{2t-k+2\alpha} {2t+1+2l-2\alpha}} & \quad \mbox{if}\quad l>0. 
 \end{array}
\right.
\end{equation}

\noindent From Equations~\ref{eq:l-two-solution} and \ref{eq:suma-harmonicos-esfericos} 
it 
is clear that the wave functions can be completely written in terms of spherical 
harmonics of both solid angles, $Y_{lm}(\Omega_1)$ and 
$Y_{lm}(\Omega_2$).

So, if $\Psi(1,2)$ is any given superposition of two-electron wave functions, the corresponding reduced
density matrix is
\begin{equation}\label{eq:rho-reducida-spherium}
 \rho(1,1')=\int\Psi^*(1,2)\Psi(1',2)d\Omega_2 .
\end{equation}

\noindent To calculate the necessary eigenvalues, the matrix representation of the density operator in Equation~\ref{eq:rho-reducida-spherium} is calculated in a basis. It is natural to use the spherical harmonics

\begin{equation}\label{eq:elements-rho-spherium}
 [\rho]_{j,s,j',s'}=\int d\Omega_1\,\int d\Omega_{1'}\, Y^*_{j,s}(1) \rho(1,1')Y_{j',s'}(1').
\end{equation}

\noindent Fortunately, to evaluate explicitly and exactly both  Equations~\ref{eq:rho-reducida-spherium} and \ref{eq:elements-rho-spherium} only integrals involving two and three spherical harmonics are required,


\begin{eqnarray}
\label{tr}
\int 
Y^*_{l_3,m_3}(\theta,\varphi)Y_{l_2,m_2}(\theta,\varphi)Y_{l_1,m_1}(\theta,
\varphi)d\Omega&=&\Big[\frac{(2l_1+1)(2l_2+1)}{4\pi(2l_3+1)}\Big]^{1/2}\times 
 \\
&&C(l_1 , m_1;l_2 ,m_2;l_3,m_3)C(l_1,
0;l_2,0;l_3,0),\nonumber
\end{eqnarray}

\noindent together with the formula

\begin{eqnarray}
\label{do}
Y_{l_1,m_1}(\theta,\varphi)Y_{l_2,m_2}(\theta,\varphi)&=&\sum_{L=0}^{\infty}\Big
[ 
\frac{(2l_1+1)(2l_2+1)}{4\pi(2L+1)}\Big]^{1/2}\times 
 \\
&& C(l_1,m_1;l_2,m_2;L,M)C(l_1, 
0;l_2,0;L,0)Y_{L,m_1+m_2}(\theta,\varphi).\nonumber
\end{eqnarray}\\


Figure~\ref{fig:petska} shows the results for the $L=2$ case. The superposition state
$\left|\psi_{\alpha}\right>$ of Equation~\ref{eq:superposition} is in this case
\begin{equation}\label{eq:lc_petska}
\left|\psi_{\alpha}\right>=\sqrt{\alpha}\left|\Psi_{L,M}\right>+\sqrt{1-\alpha}
\left|\Psi_{L,-M}\right>.
\end{equation}

\noindent As can be seen in Figure~\ref{fig:petska}, the two possible elections 
for $|M|=1,2$ render a convex and a concave entropy curve
for the superposition state $\left|\psi_{\alpha}\right\rangle$, respectively. The criterion, shown in 
Table~\ref{ta:petska}, correctly predicts the convexity in both cases.

\section{Laguerre-Gaussian one-photon states}\label{sec:lg-states}

Two-photon states can be constructed, from a theoretically point of view, just 
applying 
two creation operators to the vacuum state. From an experimental point of view, 
the most used method is the spontaneous parametric down conversion (SPDC) 
\cite{kok}. It is 
well known that the SPDC mechanism provides a couple of photons in an entangled 
state that can be used to perform different quantum information tasks. 
Nevertheless, the two-photon state depends on the mode function of the pump and
the phase matching conditions. So, to analyze a simpler case we focus in 
one-photon 
states and use the concept of single particle entanglement 
\cite{one-particle-entanglement} where one spatial degree of freedom of the photon 
wave-function is traced out. Since we are interested in states that are 
eigenstates of $L_z$ the Laguerre-Gaussian states are an obvious choice to test the criterion. They are given by \cite{kok,one-particle-entanglement}
\begin{equation}\label{eq:laguerre-gaussian}
u_{lm}^{LG}(k;\vec{r},t)=\frac{4\pi(-1)^{l+|m|}l!}{s^{2(l+|m|+1)}(z)}r^{|m|}e^{
im\phi}L_l^{|m|}\Big(\frac{r^2}{s^2(z)}\Big)\,
e^{ikz-i\omega_kt-r^2/s^2(z)},
\end{equation}

\noindent where $r,\phi,z$ are the usual cylindrical coordinates, $m$ is the 
quantum number of $z$ component of the orbital angular momentum, $L_l^{|m|}$ 
are the modified Laguerre polynomials and the waist function 
$s(z)$ is 
a classical quantity that quantifies the width of the beam along the $z$ 
direction

\begin{equation}
 s^2(z)=s_0^2+i\frac{2z}{k}.
\end{equation}

\noindent In the following we drop the superscript in $u_{lm}^{LG}$ to simplify the 
notation and
make the change of variable $x=r/s(z)$. Let us recall that the 
Laguerre-Gaussian states are transverse modes that describe the free propagation 
of a photon with energy $\hbar \omega_k$, and are solutions of the Helmholtz 
equation in the paraxial approximation.

At this stage the procedure and the quantities to be calculated are well known 
so, in 
this Section, we include which states are studied and the formal expression for 
the reduced density matrix.

For a  superposition given by

\begin{equation}\label{eq:psi_fotones}
 \left|\psi_{\alpha}\right>=\sqrt{\alpha} \left|u_{lm}\right>+\sqrt{1-\alpha}\left|u_{l'm'}\right>,
\end{equation}

\noindent where both $l,m$ and $l^{\prime},m^{\prime}$ are quantum numbers 
compatible 
with Equation~\ref{eq:laguerre-gaussian}, we calculate the reduced density 
matrix

\begin{eqnarray}
 \rho_{red}(x,x')=\int_{-\infty}^\infty \textrm{d}y\,  &&
\left\lbrace\alpha u^*_{lm}(x,y,z)\,u_{lm}(x',y,z)\right.  \nonumber \\
&&+\left(1-\alpha\right)\,u^*_{l'm'}(x,y,z)\,u_{l'm'}(x',y,
z) 
\nonumber \\ 
&& +\,\sqrt{\alpha(1-\alpha)}\, \left[u^*_{lm}(x,y,z)\,u_{l'm'}(x',y,z)\right. \nonumber \\ 
&& +\left.\left. u^{*}_{l'm'}(x,y,z)\,u_{lm}(x',y,z)\right]\,\right\rbrace,
\end{eqnarray}

\noindent where $z$ is considered as a parameter and $x,y$ are the usual 
Cartesian 
coordinates perpendicular to $z$. We consider reduced density matrices at constant values of $z$ since the LG states in Equation~\ref{eq:laguerre-gaussian} are 
not square-integrable functions. To avoid this kind of assumption it is possible 
to implement the calculation of reduced density matrices and the corresponding 
entropies using LG modes in a cavity  \cite{cavity1,cavity2}, which are 
square-integrable and very similar to those in Equation~\ref{eq:laguerre-gaussian}, 
so the dependency on the $z$ variable can be traced out completely.

To test the criterion it is necessary to obtain the matrix elements

\begin{equation}
[\rho_{red}]_{ab}=\int_{-\infty}^{\infty}dx\int_{-\infty}^{\infty}
dx'f_a(x)\rho_ {red}(x,x')f_b(x'),
\end{equation}

\noindent where the one-coordinate basis functions used are the Hermite functions
\begin{equation}
 f_a(x)=\frac{1}{\sqrt{\sqrt{\pi}2^a a!}}H_a(x)e^{-x^2/2}, \quad a=0,1,2 \ldots \;.
\end{equation}

\begin{figure}[h] 
\centering
\includegraphics[height=7cm]{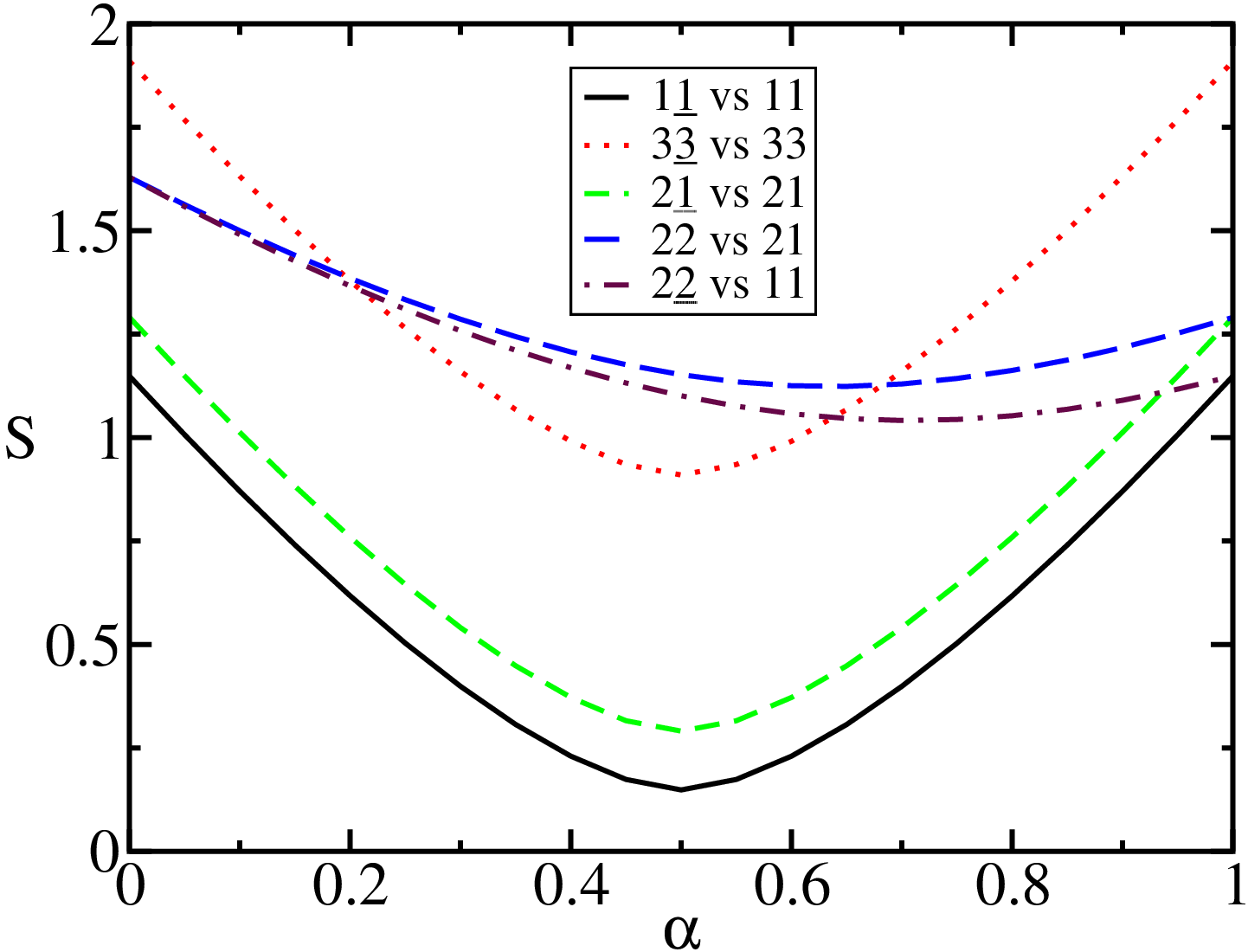} \\
\caption{\label{fig:fotones}von Neumann entropy as a function of the combination
parameter $\alpha$, for states $\left|\psi_{a}\right>$ from Equation~\ref{eq:psi_fotones}.
The symmetric combinations ($l=l'$ and $m=-m'$) $(l_{\alpha=0},m_{\alpha=0},l_{\alpha=1},m_{\alpha=1})$=$(1,-1,1,1),(3,-3,3,3)$ and $(2,-1,2,1)$ are shown with black, red pointed and green dashed curves, 
respectively. The asymmetric combinations $(l_0,m_0,l_1,m_1)$=$(2,2,2,1),(2,-2,1,1)$
are shown with blue dashed and brown dash-dotted curves respectively.}
\end{figure}

Figure \ref{fig:fotones} shows the results obtained for these states. 
As the results included in Table~\ref{ta:fotones} show, the criterion correctly 
predicts the convexity of the superpositions  considered.

The one-photon eigenstates can not provide a superposition with a concave von 
Neumann entropy because the one-photon wave function depends on the two 
transversal coordinates, $x$ and $y$,  in exactly the same way. It 
is clear from Equation~\ref{eq:laguerre-gaussian} that the state with quantum 
number $m$ depends on the same set of Laguerre polynomials than the state with 
quantum number $-m$. So, when tracing one or the other coordinate  the 
corresponding reduced density matrices that enter in the calculation of the 
criterion both ''occupy'' the same portion of the one-coordinate Hilbert space. 
As a consequence their remaining entropy always overcome their not-shared 
entropy. 

\begin{table}[h]
 \centering
\begin{tabular}{|c|c|c|c|c|c|c|}
\hline 
$lm$ & $l'm'$ & convexity &  $Q_c$ &$S_{NS}$ &$S_{R}$  \\ 
\hline 
11 & 1\underline{1} & convex & + & 0 & 0.796 \\ 
\hline 
21 & 2\underline{1} & convex & + &0 & 0.894 \\ 
\hline 
21 & 22 & convex & + & 0.194 & 0.700 \\ 
\hline 
11 & 2\underline{2} & convex & + & 0.187 & 0.608\\
\hline
33 & 3\underline{3} & convex & + & 0 & 1.323\\
\hline
\end{tabular}
\caption{\label{ta:fotones}The values of the convexity, the criterion prediction 
and 
the entropies of the curves shown
in Figure~\ref{fig:fotones}.}
\end{table}

\section{Two-particle total angular momentum eigenstates}
\label{sec:am-criterion}

So far, the examples analyzed in the previous Sections provide a strong evidence 
of 
the validity of the criterion stated in this work about the convexity 
of superpositions of degenerate states. Regrettably, all of them have reduced 
density matrices $\rho_{A,0}$ with non-degenerate eigenvalues.   

 An exact example showing reduced density matrices with degenerate eigenvalues 
can 
be constructed from the addition of two angular momentum 
operators. As usual we consider
\begin{equation}
\mathbf{L}^2 \left| L M; l_1,l_2\right\rangle = L(L+1) \left| L M; l
_1,l_2\right\rangle \quad ; \quad L_z \left| L M; l_1,l_2\right\rangle = M \left| L M; 
l_1,l_2\right\rangle,
\end{equation}

\noindent where $\mathbf{L}^2= (\mathbf{L}_1+ \mathbf{L}_2)^2$, $L_z = L_z^{(1)} + L_z^{(2)}$. For each particle the square of the angular momentum operator, $\mathbf{L}_i^2$, and its $z$ component, $L_z^{(i)}$, have common eigenfunctions which satisfy

\begin{eqnarray}
\mathbf{L}_i^2 \left|l_i,m_i\right\rangle =  l_i(l_i+1) 
\left|l_i,m_i\right\rangle \quad ; \quad L_z^i \left|l_i,m_i\right\rangle = m_i 
\left|l_i,m_i\right\rangle ,
\end{eqnarray}

\noindent for $i=1,2$.

To fix ideas, let us consider the Hamiltonian for two interacting spins given by
\begin{equation}\label{eq:hamiltonian-spins}
H = \mathbf{L}^2 - L_z^2 = \mathbf{L}_1^2 + \mathbf{L}_2^2 + 2 \mathbf{L}_1 \cdot \mathbf{L}_2 - L_z^2.
\end{equation}
\noindent and choose two spins with the same angular quantum number,  $l_1=l_2=\ell$. Consistently with the 
superpositions analyzed in previous Sections, we consider states given by

\begin{equation}\label{eq:rhoab_angular-sec}
\Phi=\sqrt{\alpha}\mathcal{Y}^{L,L}_{\ell,\ell}+\sqrt{1-\alpha}\mathcal{Y}^{L,-L}_{\ell,
\ell},
\end{equation}

\noindent where

\begin{equation}
\label{eq:clebsch-gordan}
\mathcal{Y}_{\ell,\ell}^{L,\pm 
L}=\sum_{m_1,m_2}C(\ell,m_1;\ell,m_2;L,\pm L)Y_{\ell,m_1}
(\Omega_1)Y_{\ell,m_2}(\Omega_2),
\end{equation}

\noindent and $C(\ell,m_1;\ell,m_2;L,M)$ are the Clebsch-Gordan coefficients.

The states $\mathcal{Y}^{L,L}_{\ell,\ell}$ and $\mathcal{Y}^{L,-L}_{\ell,\ell}$ are 
degenerate since

\begin{equation}
 H \mathcal{Y}^{L,\pm L}_{\ell,\ell} = L \mathcal{Y}^{L,\pm L}_{\ell,\ell} ,
\end{equation}

\noindent and for fixed $L$ they are the states with minimum energy, besides $0\leq L \leq 2\ell$.

We include in Appendix \ref{ap:angular-momentum} the necessary algebraic 
details 
to evaluate explicitly and exactly the entries of the different reduced density 
matrices and their eigenvalues. 


\begin{figure}
    \centering
    \includegraphics[height=6cm]{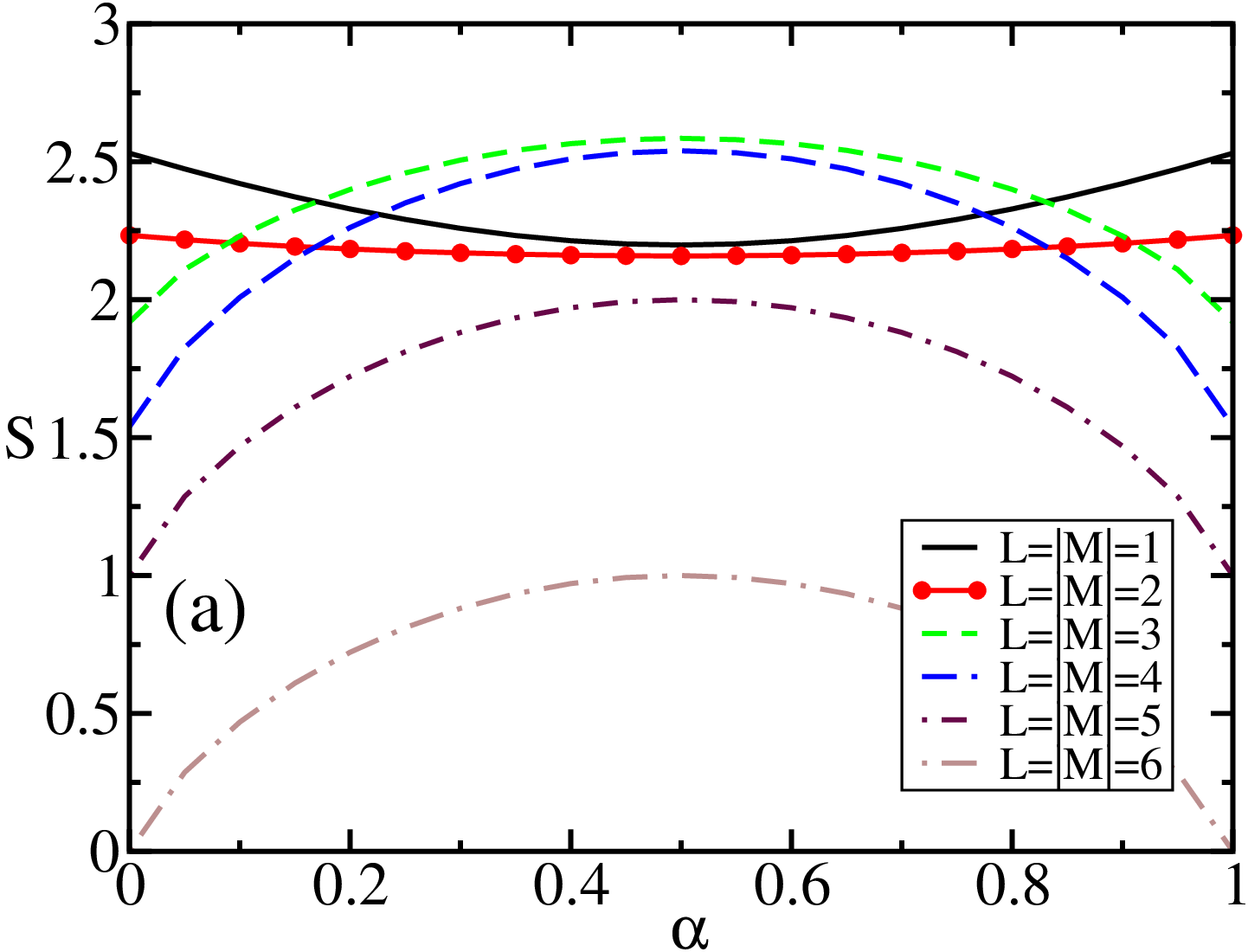}       
    \includegraphics[height=6cm]{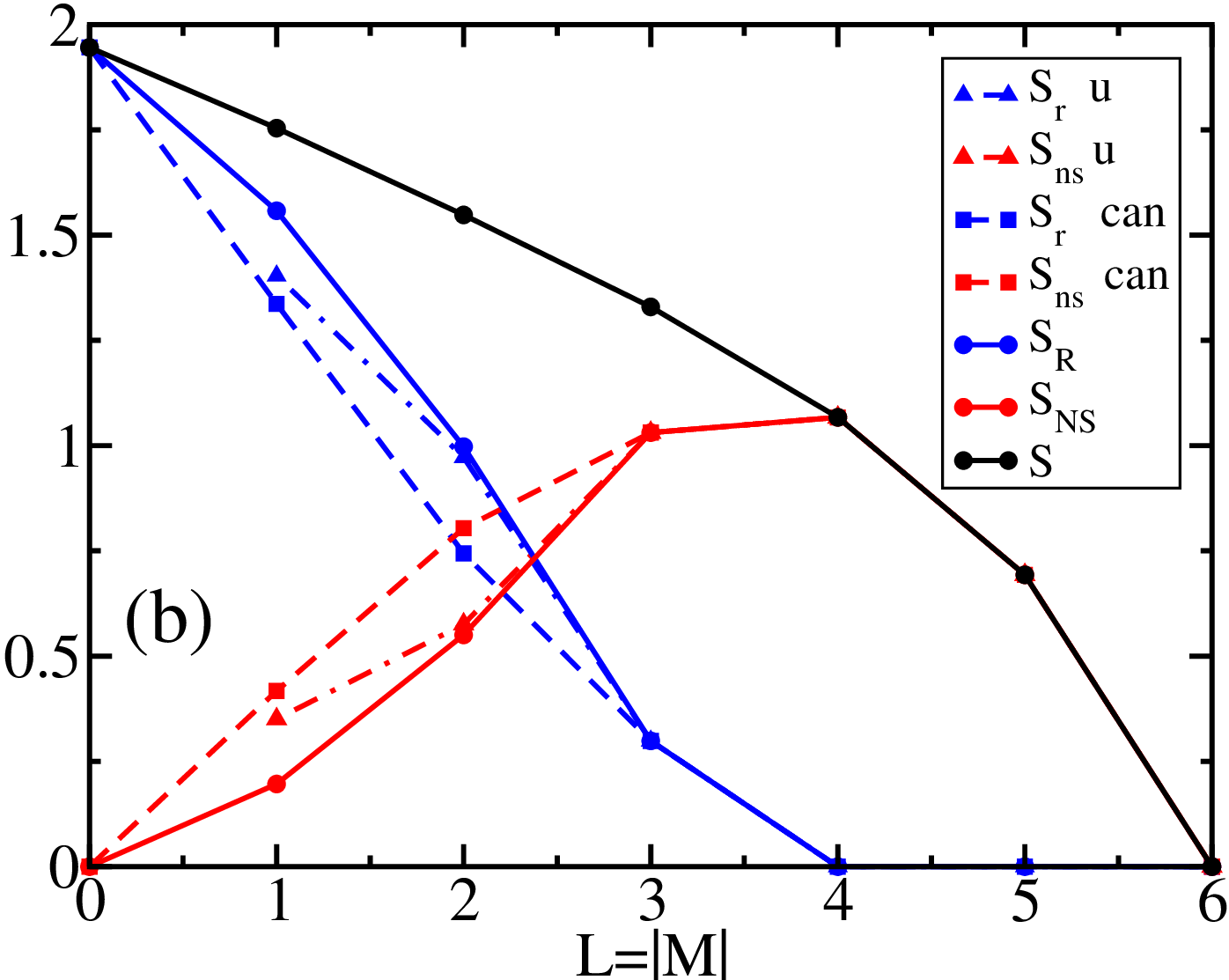}
    \caption{\label{fig:angular-example}
    (a) von Neumann entropy as a function of the combination parameter
    $\alpha$, for the states of Equation~\ref{eq:rhoab_angular-sec} with
$\ell=3$ and $L=|M|$. $|M|$ and $L$ 
    values for each curve are given in the legend. (b) The von 
Neumann, the remaining and not-shared entropies calculated for the cases shown 
in (a) are shown using black, red and blue solid circular dots, 
respectively. The square dots correspond to the not-shareable entropy 
calculated using the canonical angular momentum basis and the triangular ones 
to the entropy calculated using other basis (see the text for details).}
\end{figure}

\begin{table}[h]
 \centering\begin{tabular}{|c|c|c|c|c|c|}
\hline 
$L$ & $|M|$ & convexity & $Q_c$ &$S_{NS}$ & $S_{R}$ \\ 
\hline 
1 & 1 & convex & + & 0.196 & 1.558 \\ 
\hline
2 & 2 & convex & + & 0.550 & 0.997  \\ 
\hline
3 & 3 & concave & -- & 1.031 & 0.299 \\ 
\hline
4 & 4 & concave & -- & 1.067 & 0\\ 
\hline
5 & 5 & concave & -- & 0.693 &0 \\ 
\hline
6 & 6 & concave & 0 &0 & 0\\ 
\hline
\end{tabular}
\caption{\label{ta:angular} The values of the convexity, the criterion prediction 
and the entropies of the curves shown in Figure~\ref{fig:angular-example}.}
\end{table}

Figure~\ref{fig:angular-example}(a) shows the explicit evaluation of 
$S(\rho_{A,\alpha})$ for the case $\ell=3$, $|M|=L$ and $L=1,2,3,4,5$ and $6$. 
It can be appreciated that the cases $L=3,4,5$ and $6$ have a different 
convexity than the $L=1$ and $2$ cases. On the other hand, the solid circular 
dots in Figure~\ref{fig:angular-example}(b) correspond to the values of the 
{\em not-shared} and {\em remaining} entropies calculated using  
Equations~\ref{eq:not-shared} and \ref{eq:remaining-entropy}. The lines are 
included as a guide to the eye. It is clear that using these entropies the 
criterion   detects the correct convexity of all the cases, as it is shown in 
Table~\ref{ta:angular}.

It is interesting to analyze in some detail the case $\ell=3$ and $L=2$. The reduced density 
matrices can be calculated explicitly and, in the standard one-particle angular 
momentum basis $\left\lbrace |m=3\rangle, |m=2\rangle, \ldots, |m=-3\rangle  
\right\rbrace$, they are diagonal matrices
\begin{equation}\label{eq:rhoa-l3}
\rho_{A,0} = diag\left\lbrace \frac{5}{42},\frac{5}{21},\frac{2}{7},\frac{5}{21},\frac{5}{42},0,0 \right\rbrace,
\end{equation}
and 

\begin{equation}\label{eq:rhob-l3}
\rho_{A,1} = diag\left\lbrace 0,0, \frac{5}{42},\frac{5}{21},\frac{2}{7},\frac{5}{21},\frac{5}{42} \right\rbrace .
\end{equation}

It is clear that the matrix $\rho_{A,0}$ has two pairs of degenerate 
eigenvalues, the 
first and fifth one are the first pair ($\lambda=5/42$), and the second and 
fourth are the second one ($\lambda=5/21$). Accordingly with 
Equation~\ref{eq:tilde-entropy}, to contribute to the not-shared entropy we must 
compare two times the eigenvalue of $\rho_{A,0}$ with the sum of eigenvalues of 
$\rho_{A,1}$ that lie in the same sub-space. But since $2\times\frac{5}{42}=\frac{5}{6}\times \frac{2}{7}< 
\frac{2}{7}$,  
\begin{equation}
\min \left\lbrace \sum_{i=1}^2 \Theta\left[\lambda_{\nu}^{A,0} - 
\langle \rho_{A,1}\rangle_{i,\nu}\right] \log_{2}\left(\frac{1}{\lambda_{\nu}^{A,0}}\right) 
\right\rbrace= 0.
\end{equation}

On the other hand, the term associated to the eigenvalue $5/21$ contributes with
\begin{equation}
\tilde{S} = \left( 2\times \frac{5}{21} - \frac{5}{21}\right) \log_{2}\frac{21}{5} .
\end{equation}
Collecting these results, the not-shared entropy is equal to

\begin{equation}
S_{NS} = \left(\frac{2}{7} - \frac{5}{42}\right) \log_{2}\frac{7}{2} + 
\left(\frac{5}{21} \right) \log_{2}\frac{21}{5}.
\end{equation}

It is instructive, and simple to do, to check what happens if the minimization 
implied in the definition of the not-shared entropy is not performed, {\em i.e.} 
what values are obtained for different choices of the one-dimensional projectors 
associated to the degenerate eigenvalues. Choosing the canonical one-particle 
angular momentum basis to generate the one-dimensional projectors gives the  
not-shareable entropy
\begin{equation}\label{eq:not-shareable-canonical}
S_{ns} = -\sum_{\lambda_i^{A,0}\neq 0}^7 \Theta\left[ \lambda_i^{A,0} - 
\lambda_i^{A,1}\right] \log_{2}\left(\lambda_i^{A,0}\right), 
\end{equation}

\noindent where the eigenvalues, for the case $L=2$, are those in Equations~\ref{eq:rhoa-l3} and  
\ref{eq:rhob-l3}. The values obtained using Equation~\ref{eq:not-shareable-canonical} 
are shown in Figure~\ref{fig:angular-example}(b) as square solid blue dots. Not 
surprisingly, the values are larger than those of $S_{NS}$, but more 
interestingly, it is clear that for $L=2$ the not-shareable entropy is larger 
than the remaining one, which could lead to an incorrect assessment of the 
convexity. Sometimes, choosing a particular basis to obtain the projectors 
$P_i^{A,0}$ could give very good results for particular values of $\ell$ and 
$|M|=L$, for instance, in Figure~\ref{fig:angular-example} the values obtained 
using a particular basis are shown using triangular dots. This election provides 
value of the not-shareable entropy larger than those of $S_{NS}$ and predicts 
correctly the curvature.

We have tested a very large number of cases, up to $\ell=12$, which can be 
done quite fast and efficiently given the simplicity of the bipartite states 
$\mathcal{Y}_{\ell,\ell}^{L,M}$ and for all these cases the criterion predicts 
correctly the curvature of the superposition of states.


\section{Discussion and conclusions}\label{sec:discussion}

From a theoretical point of view, the amount of analytical work involved in 
the 
examples presented, the two two-dimensional  harmonic oscillators and 
the spherium, indicates how difficult it is to construct exact cases to test the 
convexity criterion. All the quantities involved, in particular the matrix 
elements of the reduced density matrices, involve a large number of nested sums, 
so its evaluation time grows as $M^{nes+1}$, where $nes$ is the number of 
nested 
sums and $M$ is the largest one-particle basis set size that it is necessary to 
use in order to guarantee  the normalization of the reduced density matrix. 

The criterion could be tested using pure states of many-body 
models (spin chains) which, in some cases, have exact solutions. The reduced 
matrices can be obtained using the adequate spin correlation functions for 
small subsystems. Since so far we have studied only bipartite systems made of 
two susbsystems whose Hilbert spaces have the same dimension it is not clear 
if some amendments are in order for the criterion to work in the spin chain 
setting. Work along this line is in progress.

Our results imply that,  most likely, there should be a theorem about the 
convexity of the von 
Neumann entropy of superpositions of pure states but, so far, we have not been 
able to formulate the precise hypotheses that make it work, {\em i.e.} we know 
that the superpositions of pure degenerate eigenstates satisfy the requirements 
to have a defined convexity but we do not have an algorithm that allows us to 
generate a number of 
eigenvalues, eigenfunctions (or projectors) to construct $\rho_{A,0}$ 
and $\rho_{A,1}$ 
and $\rho_{A,\alpha}$ and guarantee that $S(\rho_{A,\alpha})$ will be convex (or 
concave). In this 
sense, the conditions that the superposition is made of two degenerate states 
with  quantum number $\pm m$, where $m$ is the quantum number associate to the 
$z$-component of the total angular momentum, seem to be sufficient 
for states defined over hyper-spheres or that have the same asymptotic behavior 
that the  harmonic oscillator eigenfunctions. 

In the same sense that in the paragraph above, it is not necessary that 
$S(\rho_{A,0})=S(\rho_{A,1})$ to ensure that the von Neumann entropy, 
$S(\rho_{A,\alpha})$ has a well defined convexity (concavity). Nevertheless, it 
is worth to point out that if $\rho_{A,0} = \mbox{Tr}_{B}(|\psi_L\rangle \langle 
\psi_L|)$ or $\rho_{A,0} = \mbox{Tr}_{B}(|\psi_{-L}\rangle \langle \psi_{-L}|)$ the 
criterion predicts exactly the same convexity (concavity). In other words, the 
criterion holds even when  $\rho_{A,0}$ and $\rho_{A,1}$ are not isospectral.

In contradistinction to what happens to the entropy of one-photon states, for the eigenstates of the total angular momentum it is possible to find concave and convex functions. For example,  for the states $|2\ell,2\ell,\ell,\ell\rangle$ and 
$|2\ell,-2\ell,\ell\ell\rangle$, once a 
particle is traced out, they result in orthogonal states that do not share entropy 
 and, consequently, the von Neumann entropy of the superposition is concave. 
Besides, for some value $k$ the states $ 
\mbox{Tr}_{B}(|2\ell-k,2\ell-k,\ell,\ell\rangle \langle 
2\ell-k,2\ell-k,\ell,\ell|)$ 
and $ 
\mbox{Tr}_{B}(|2\ell-k,-(2\ell-k),\ell,\ell\rangle \langle 
2\ell-k,-(2\ell-k),\ell,\ell|)$ are not longer orthogonal and share some 
entropy. At some larger value  $k_c>k$ the remaining entropy overcomes the 
not-shared and the von Neumann entropy $S(\rho_{A,\alpha})$ becomes convex.

In Reference~\cite{Garagiola2016} it was envisaged that certain superpositions 
of degenerate bipartite 
states could have definite convexity (concavity) and that the extremal states 
would 
also be  eigenstates of other observable of the system \cite{Garagiola2016}, it 
was 
required that the system under study had at least two conserved quantities. In 
this work 
we have restricted ourselves to the case where those quantities are the energy 
and the $z$ component of the total angular momentum, which is very reasonable 
for systems with a preferred direction. For particle systems it is difficult to 
construct other conserved quantities beyond the Hamiltonian, the total angular 
momentum or some of its components, unless that some superintegrable system is 
considered. There are some examples of  two- and three-body 
superintegrable problems in dimensions two and three, where the conserved 
quantities are polynomials of the momentum operator Cartesian components. 
Currently, we are  studying  the convexity properties of superpositions of 
degenerate states in this kind of problems.  

{
As a final comment, we want to return to the a subject  that we raised at the 
end of  Section~\ref{sec:criterion}, where we stated that it could be desirable 
to 
formulate the not-shared entropy, Equation~\ref{eq:not-shared}, without 
resorting to the spectral decompositions of the reduced density operators. 
Here, we discuss some numerical tests that we implemented on the examples 
considered in Section~\ref{sec:am-criterion}, {\em i.e.} orbital angular 
momentum states with quantum number $L$. Consider  a set of 
one-dimensional projectors 
$\lbrace P_{\alpha}\rbrace_{\alpha=1}^{2L+1}$ that are mutually orthogonal 
and 
such that $\sum_{\alpha} P_{\alpha} = I$. Besides, consider the quantity
\[
 \tilde{S} = -\sum_{\alpha}  \max\left[\langle\rho_{A,0} \rangle_{\alpha}  
-\langle\rho_{A,1} \rangle_{\alpha} , 0 \right] \log\left(  \langle\rho_{A,0} 
\rangle_{\alpha} \right),
\]
where $\langle\rho_{A,i} \rangle_{\alpha} = \mbox{Tr}( \rho_{A,i} P_{\alpha})$. 
For very small values of $L$ it is numerically feasible to show that
\begin{equation}\label{eq:measure-criterion}
 min (S - 2 \tilde{S}) = S - 2 S_{NS}, 
\end{equation}
where the minimum was obtained  generating randomly families of orthogonal 
projectors and evaluating $\tilde{S}$. For larger values of $L$ the 
number of random families of projectors necessary to pick up approximately the 
value of the minimum grows so  fast that a more educated sampling becomes 
mandatory. Choosing random sets of projectors close enough to the 
eigen-projectors of $\rho_{A,0}$   
Equation~\ref{eq:measure-criterion} was  verified for moderate values of $L$ 
as the ones studied in Section~\ref{sec:am-criterion}. Note that if for 
a set of projectors $ S - 2 \tilde{S} <0$ then this is sufficient to affirm 
that the superposition will be concave. Further work along these lines is under 
progress.
}

\section*{Acknowledgements}

We acknowledge CONICET (PIP-11220150100327CO) for partial financial support. N.G. and O.O. also acknowledges SECYT-UNC for partial financial support.

\appendix

\section{Angular momentum reduced density matrices example} \label{ap:angular-momentum}

The simpler example that allow to construct bipartite states to test if the von 
Neumann entropy of a given superposition $S(\rho_{A,\alpha})$ is a convex 
function can be constructed from the states

\begin{equation}
\label{ang}
\mathcal{Y}_{l_1,l_2}^{L,M}=\sum_{m_1,m_2}C(l_1,m_1;l_2,m_2;L,M)Y_{l_1,m_1}
(\Omega_1)Y_{l_2,m_2}(\Omega_2),
\end{equation}

\noindent where $C(l_1,m_1;l_2,m_2;L,M)$ are the Clebsch-Gordan coefficients and 
$Y_{l_1,m_1}$ are 
the usual spherical harmonics. For superpositions of the form

\begin{equation}\label{eq:rhoab_angular}
\Phi=\sqrt{\alpha}\mathcal{Y}^{L,M}_{l_1,l_2}+\sqrt{1-\alpha}\mathcal{Y}^{L,-M}_{l_1,
l_2},
\end{equation}

\noindent the reduced density operator is

\begin{widetext}

\begin{eqnarray}\label{eq:rho-red-alpha}
\rho_{red}^{\alpha}(\Omega_1,\Omega_1')&=&\int 
\Phi^*(\Omega_1,\Omega_2)\Phi(\Omega_1',\Omega_2) d\Omega_2 \\
\nonumber \\
&=& \int \Big( \alpha 
\mathcal{Y}_{l_1,l_2}^{*L,M}(\Omega_1,\Omega_2)\mathcal{Y}_{l_1,l_2}^{L,M}
(\Omega_1',\Omega_2) + \nonumber \\
\nonumber \\
& &(1-\alpha)\mathcal{Y}_{l_1,l_2}^{*L,-M}(\Omega_1,\Omega_2)\mathcal{Y}_{l_1,l_2} 
^{L,-M}(\Omega_1',\Omega_2) + \nonumber \\
\nonumber \\
& &\sqrt{\alpha(1-\alpha)}\Big[\mathcal{Y}_{l_1,l_2}^{*L,M}(\Omega_1,
\Omega_2)\mathcal{Y}_{l_1,l_2}^{L, -M } (\Omega_1',\Omega_2)
+\nonumber\\
\nonumber \\
& &\mathcal{Y}_{l_1,l_2}^{*L,-M}(\Omega_1,\Omega_2)\mathcal{Y}_{l_1,l_2
} ^ { L , M } (\Omega_1',\Omega_2)\Big ]\Big)d\Omega_2 \nonumber .
\end{eqnarray}

The reduced density matrix above is function of both solid angles $\Omega_1$ 
and 
$\Omega_1^{\prime}$, so it is logical to look for its expression  in the 
spherical harmonics basis, where its elements are given by

\begin{equation}
\left[\rho_{red}^{\alpha}\right]_{i,j}=\int 
Y_{l_1,i}(\Omega_1)\rho_{red}^{\alpha}(\Omega_1,\Omega_1')Y^*_{l_1,j}
(\Omega_1')d\Omega_1d\Omega_1' .
\end{equation}

After some tedious, but straightforward algebra, it can be shown that

\begin{eqnarray}
\left[\rho_{red}^{\alpha}\right]_{i,j}=\sum_{m_2}\Big(\alpha C(l_1,i;l_2,m_2;L,M)C(l_1,j;l_2,m_2;L,
M)+ \\
\nonumber \\
(1-\alpha)C(l_1,i;l_2,m_2;L,-M)C(l_1,j;l_2,m_2;L,-M)+\nonumber \\
\nonumber \\
\sqrt{\alpha(1-\alpha)}\Big[C(l_1,i;l_2,m_2;L,M)C(l_1,j;l_2,m_2;L,
-M)+\nonumber \\
\nonumber \\
C(l_1 , i;l_2 , m_2;L,-M)C(l_1,j;l_2,m_2;L,M)\Big]\Big)\nonumber ,
\end{eqnarray}

\noindent where $|i|,|j| \leq l_1$. 

\end{widetext}

\section{Two harmonic oscillators related expressions} \label{ap:two-oscillators}

In this Appendix we present explicit expressions of the two harmonic oscillator states
that allow the computation of the density matrix eigenvalues. The angular momentum two-particle eigenstates
in Equations 
\ref{eq:two-oscillators-eigenvalues} 
and \ref{eq:two-oscillators-angular-momentum}, can be written as

\begin{equation}\label{eq:two-interacting-osc-pol}
\left| n,m,l,p\right\rangle = |n,m\rangle_{R,\phi_R} |l,p\rangle_{r,\phi_r} ,
\end{equation}
where the sub-indexes indicate for which oscillator the vector sate is an eigenstate, following the convention that $R$ designates the centered coordinates oscillator and $r$ the relative coordinates one.  

Each one of the one-particle angular momentum  eigenvectors in Equation \ref{eq:two-interacting-osc-pol}
can be written as linear combination of Cartesian oscillator states, 
$\left|n_x,n_y\right\rangle^c = \left|n_x\right\rangle^c 
\left|n_y\right\rangle^c$~\cite{Cohen}, as follows

\begin{eqnarray}\label{eq:one-particle-cylindical}
\left|n,m\right\rangle_{\tilde{r},\tilde{\phi}}&=&\frac{1}{\sqrt{n!(n+|m|)!}2^{(2n+|m|)/2}} \times \\ \nonumber
&& \sum_{j=0}^n\sum_{k=0}^{n+|m|}{n\choose j}
{n+|m| \choose k} 
i^{sgn(m)(k-j)}\times \\ \nonumber
&& \sqrt{(2n+|m|-j-k)!(j+k)!}\left|2n+|m|-j-k,j+k\right\rangle^c_{\tilde{x},\tilde{y}},
\end{eqnarray}

\noindent where $\tilde{r},\tilde{\phi}$ stands for $R,\phi_R$ or $r,\phi_r$, and the same convention holds for the Cartesian coordinates, $sgn(m)=1$ for $m \geq 0$ and  $sgn(m)=-1$ for $m<0$.  The 
one-particle Cartesian eigenfunctions corresponding to the vector state $\left|n_{\tilde{x}},n_{\tilde{y}}\right\rangle^c = 
|n_{\tilde{x}}\rangle^c |n_{\tilde{y}}\rangle^c$ are given by

\begin{eqnarray}\label{eq:one-particle-cartesian}
\langle \tilde{x}, \tilde{y} \left|n_{\tilde{x}},n_{\tilde{y}}\right\rangle^c  &=&\sqrt{\frac{\omega_{\tilde{r}}}{\pi2^{n_{\tilde{x}}+n_{\tilde{y}}+1}n_{\tilde{x}}!n_{\tilde{y}}!}}e^{
-\omega_{\tilde{r}}(\tilde{x}^2+\tilde{y}^2)/4}H_{n_{\tilde{x}}}\left(\sqrt{\frac{\omega_{\tilde{r}}}{2}} \tilde{x}\right)H_{n_{\tilde{y}}}\left(
\sqrt{\frac{\omega_{\tilde{r}}}{2}}\tilde{y}\right), 
\end{eqnarray}

\noindent where $H_n$ is the Hermite polynomial 
of $n-$th degree, with $n=0,1,2. \ldots$ The energy of these two-dimensional 
harmonic 
oscillator states 
is $E_{n_{\tilde{x}},n_{\tilde{y}}}=\omega_{\tilde{r}}(n_{\tilde{x}}+n_{\tilde{y}}+1)$.

Collecting the results above, we get that

\begin{eqnarray}
 \left| n,m,l,p\right\rangle &=& \sum_{j=0}^{n} \sum_{k=0}^{n+|m|} 
 \sum_{r=0}^{l} \sum_{s=0}^{l+|p|}
\kappa(n,m,j,k)  
 \kappa(l,p,r,s)|2n+|m|-j-k\rangle^c_{x_1+x_2} 
\times \nonumber \\ 
 && |j+k\rangle^c_{y_1+y_2}
 |2l+|p|-r-s\rangle^c_{x_1-x_2} |r+s\rangle^c_{y_1-y_2},
\end{eqnarray}

\noindent the notation indicates which oscillator and  coordinates must be used 
to obtain 
the corresponding eigenfunctions

\begin{eqnarray}\label{eq:wave-function-separated-coordinates}
\psi_{n,l,m,p}(x_1,y_1,x_2,y_2) &=& \sum_{j=0}^{n} \sum_{k=0}^{n+|m|} 
\sum_{r=0}^{l} \sum_{s=0}^{l+|p|}
\kappa(n,m,j,k) 
 \kappa(l,p,r,s) f_{2n+|m|-j-k}^{\omega_R}\left(\frac{x_1+x_2}{\sqrt{2}}\right) \times 
\nonumber \\ 
&& 
f_{j+k}^{\omega_R}\left(\frac{y_1+y_2}{\sqrt{2}}\right) 
f_{2l+|p|-r-s}^{\omega_r}\left(\frac{x_1-x_2}{\sqrt{2}}\right) 
f_{r+s}^{\omega_r}\left(\frac{y_1-y_2}{\sqrt{2}}\right) \nonumber\\
 &=& \sum_{j=0}^{n} \sum_{k=0}^{n+|m|} 
\sum_{r=0}^{l}
\sum_{s=0}^{l+|p|} \kappa(n,m,j,k)   \kappa(l,p,r,s) \times \nonumber \\ 
&&\Upsilon_{2n+|m|-j-k\,,\,j+k\,,\,2l+|p|-r-s\,,\, r+s }(x_1,y_1,x_2,y_2).
\end{eqnarray}

\noindent where $f_n^{\omega}(x)$ are the eigenfunctions of a one dimensional harmonic oscillator
of frequency 
$\omega$ and quantum number $n$ and the last equation defines $\Upsilon_{a,b,c,d}$. 

The  one-particle reduced density matrix for an eigenfunction $\psi_{n,l,m,p}$ 
can 
be obtained exactly from
\begin{equation}
\rho_A(\vec{x}_1,\vec{x}_1^{\prime}) = \int d\vec{x}_2 
\psi^*_{n,m,l,p}(\vec{x}_1,\vec{x}_2) 
\psi_{n,m,l,p}(\vec{x}_1^{\prime},\vec{x}_2) ,
\end{equation}

\noindent where $\psi_{n,m,l,p}(\vec{x}_1,\vec{x}_2)$ is the two-particle 
wave function $\left|n,m,l,p\right\rangle$ written in terms of the original 
particle coordinates, Equation~\ref{eq:wave-function-separated-coordinates}. Actually, 
to implement the calculation of the  eigenvalues, $\lambda_i^A$, required to obtain the 
different entropies that enter in the convexity criterion, 
Equations~\ref{eq:not-shared}, \ref{eq:remaining-entropy}, \ref{eq:criterion1} 
and 
\ref{eq:criterion2}, it is useful to calculate the matrix elements of 
$\psi_{n,m,l,p}$ in a one-particle one-coordinate basis functions 
$\phi_i(z)$, where $z$ stands for $x$ or $y$, {\em i.e.} we first calculate

\begin{equation}\label{eq:actual-matrix-two-oscillator}
K^{n,m,l,p}_{i_1,j_1,i_2,j_2} = \int dx_1 dy_1 dx_1^{\prime} dy_1^{\prime} 
\phi_{i_1}(x_1) \phi_{j_1}(y_1) \psi_{n,m,l,p}(x_1,y_1,x_1^{\prime}, 
y_1^{\prime})  \phi_{i_2}(x_1^{\prime}) \phi_{j_2}(y_1^{\prime}) ,
\end{equation}

\noindent and then solve the eigenvalue problem

\begin{equation}\label{eq:kernel-matrix-representation}
\mathbf{K} \mathbf{u}_i = k_i^A \mathbf{u} ,
\end{equation}

\noindent where $\mathbf{K}$ is the matrix whose entries are given by 
Equation~\ref{eq:actual-matrix-two-oscillator}. Since $\psi_{n,m,l,p}$ is a 
symmetric kernel (under particle exchange), the eigenvalues of the reduced density matrix satisfy that 
\cite{symmetric-kernel}

\begin{equation}
    \lambda_i^A = (k_i^A)^2 .
\end{equation}

The algebra involved in the calculation of  the elements $K^{n,m,l,p}_{i_1,j_1,i_2,j_2}$ is 
rather cumbersome, but direct, so we write them explicitly. We compute the 
expressions for the wave function of two non-interacting harmonic oscillators, $\lambda=0$, 
and also the elements of its reduced density matrix, 
Equation~\ref{eq:actual-matrix-two-oscillator}. 
The  corresponding quantities for $\lambda\neq 0$  can be obtained following 
a completely equivalent procedure. 

Then, using the expansion of the Hermite polynomials \cite{Hermite}, we get that
\begin{eqnarray}
\Upsilon_{n,m,l,p}(x_1,y_1,x_2,y_2)&=&\sqrt{\frac{n!m!l!p!}{\pi^22^{n+m+l+p}}} 
\sum_{a=0}^{\lfloor n/2 \rfloor}\sum_{b=0}^{\lfloor m/2 
\rfloor}\sum_{c=0}^{\lfloor l/2 \rfloor}\sum_{d=0}^{\lfloor p/2 
\rfloor}
\sum_{s=0}^{n-2a}\sum_{t=0}^{m-2b}\sum_{v=0}^{l-2c}\sum_{w=0}^{p-2d}{
n-2a\choose s}{m-2b\choose t}  \nonumber \\ 
&& 
{l-2c\choose v}{p-2d\choose 
w}\frac{(-1)^{a+b+c+d+v+w}\sqrt{2}^{n-2a+m-2b+l-2c+p-2d}}{
a!b!c!d!(n-2a)!(m-2b)!(l
-2c)!(p-2d)!} \nonumber  \\ 
&&
x_2^{n-2a-s+l-2c-v}y_1^{m-2b-t+p-2d-w}
x_1^{s+v}y_2^{t+w} e^{-(x_1^2+y_1^2+x_2^2+y_2^2)/2} .
\end{eqnarray}

Now, for each $\Upsilon$ function we construct a kernel ${K_\Upsilon}$ by means of
Equation~\ref{eq:actual-matrix-two-oscillator}, 
and using the one-particle basis functions $\phi_k=f^{1}_k(x)$, we get the matrix-representation of each kernel as

\begin{eqnarray}
\left[K_{\Upsilon_{n,m,l,p}}\right]_{i_1,j_1,i_2,j_2}&=&
\sqrt{\frac{n!m!l!p!i_1!j_1!i_2!j_2!}{\pi^42^{n+m+l+p+i_1+j_
1+i_2+j_2} } } 
\sum_{a=0}^{\lfloor n/2 \rfloor}\sum_{b=0}^{\lfloor m/2 
\rfloor}\sum_{c=0}^{\lfloor l/2 \rfloor}\sum_{d=0}^{\lfloor p/2 
\rfloor}\sum_{s=0}^{n-2a}\sum_{t=0}^{m-2b}\sum_{v=0}^{l-2c}\sum_{w=0}^{p-2d}\\
\nonumber \\
&&\sum_{q=0}^{\lfloor i_1/2 \rfloor}\sum_{z=0}^{\lfloor j_1/2 
\rfloor}\sum_{g=0}^{\lfloor i_2/2 \rfloor}\sum_{e=0}^{\lfloor j_2/2 
\rfloor}
{n-2a\choose s}{m-2b\choose t}{l-2c\choose v}{p-2d\choose w}\nonumber \\
\nonumber \\
&&\frac{(-1)^{a+b+c+d+v+w+q+z+g+e}\sqrt{2}^{n-2a+m-2b+l-2c+p-2d}2^{i_1-2q+j_1-2z+i
_2-2g+j_2-2e} } { a!b!c!d!(n-2a)!(m-2b)!(l-2c)!(p-2d)!(i_1-2q)!(j_1-2z)!(i
_2-2g)!(j_2-2e)! }\nonumber \\
\nonumber \\
&&\int_{-\infty}^{\infty}x_1^{n-2a-s+l-2c-v+i_1-2q}e^{-x_1^2}dx_1 
\int_{-\infty}^{\infty}y_1^{m-2b-t+p-2d-w+j_1-2z}e^{-y_1^2}dy_1 \nonumber \\
\nonumber \\
&&\int_{-\infty}^{\infty}x_2^{s+v+i_2-2g}e^{-x_2^2}dx_2 
\int_{-\infty}^{\infty}y_2^{t+w+j_2-2e}e^{-y_2^2}dy_2 \nonumber .
\end{eqnarray}

All the integrals that appear in the last expression are obtained in terms of 
the 
Gamma function \cite{Gamma}. Finally, the matrix representation of $\mathbf{K}$ 
in 
Equation~\ref{eq:kernel-matrix-representation} can be written as

\begin{eqnarray}
\left[\mathbf{K}\right]_{i_1,j_1,i_2,j_2} &=& \sum_{j=0}^{n} \sum_{k=0}^{n+|m|} 
\sum_{r=0}^{l} 
\sum_{s=0}^{l+|p|} \kappa(n,m,j,k)   \kappa(l,p,r,s) \times \nonumber \\ 
&& \left[K_{\Upsilon_{2n+|m|-j-k,j+k,2l+|p|-r-s, r+s 
}}\right]_{i_1,j_1,i_2,j_2}    .
\end{eqnarray}

\end{document}